\documentclass[prd,superscriptaddress,twocolumn,aps,showpacs,amsmath,amssymb,floatfix]{revtex4-1}

\usepackage{color}
\usepackage{bm}
\usepackage{hyperref}
\usepackage{graphicx}
\usepackage{braket}
\usepackage{MnSymbol}
\usepackage{amssymb}
\hypersetup{colorlinks=true}
\usepackage{dcolumn}
\usepackage{amsmath}
\usepackage{amsfonts}
\usepackage{bm}
\usepackage{color}

\usepackage{hyperref}
\hypersetup{colorlinks,linkcolor=blue,urlcolor=blue,citecolor=blue}

\renewcommand{\v}[1]{\textbf{#1}}

\begin{document}
\title{{Tunneling spectroscopy of quantum spin liquids}}
\author{Elio~J.~K\"onig}
\affiliation{Department of Physics and Astronomy, Center for Materials Theory, Rutgers University, Piscataway, NJ 08854, USA}
\author{Mallika~T.~Randeria}
\affiliation{Department of Physics, Massachusetts Institute of Technology, Cambridge, Massachusetts 02139, USA}
\author{Berthold~J\"ack}
\affiliation{Princeton University, Joseph Henry Laboratory at the Department of Physics, Princeton, NJ 08544, USA}
\date{\today }

\begin{abstract}
We examine the spectroscopic signatures of tunneling through a
Kitaev quantum spin liquid (QSL) barrier in a number of experimentally relevant geometries.  We combine
contributions from elastic and inelastic tunneling processes and find that spin-flip scattering at the itinerant spinon modes gives rise to a gapped contribution to
the tunneling conductance spectrum. We address the spectral modifications that arise in a magnetic field necessary to drive the candidate material $\alpha$-RuCl$_3$ into a 
QSL phase, and we propose a lateral 1D tunnel junction as a viable setup in this regime. The characteristic
spin gap is an unambiguous signature of the fractionalized QSL excitations, distinguishing it from magnons or phonons. The results of our analysis are generically applicable to a wide variety of topological QSL systems. 
\end{abstract}

\maketitle

\textit{\textcolor{blue}{Introduction.}} Geometric frustration of localized electron spins can suppress magnetic order and favor the formation of a quantum spin liquid (QSL) state, which is characterized by a macroscopic ground state of entangled quantum spins with absent long-range order
\cite{anderson1973resonating}. The spin degree of freedom of a QSL state can fractionalize into a set of anyonic excitations, where the exactly solvable Kitaev model on a honeycomb lattice predicts the emergence of localized, gapped $\mathbb Z_2$ fluxes and an 
itinerant, relativistic Majorana spinon mode \cite{Kitaev2006}. When time reversal symmetry is broken, e.g. by the application of a magnetic field, the bulk spinon spectrum is further expected to acquire a topological mass gap, giving rise to emergent 1D chiral Majorana edge modes. Their non-Abelian quantum statistics could present avenues for implementing topologically protected quantum computation \cite{nayak2008non}, whose prospect promotes today's intense research efforts on this topic \cite{jackeli2009mott, SavaryBalents2016,zhou2017quantum,knolle2019field}.

\begin{figure}[h]
    \centering
  \includegraphics[width = .45\textwidth]{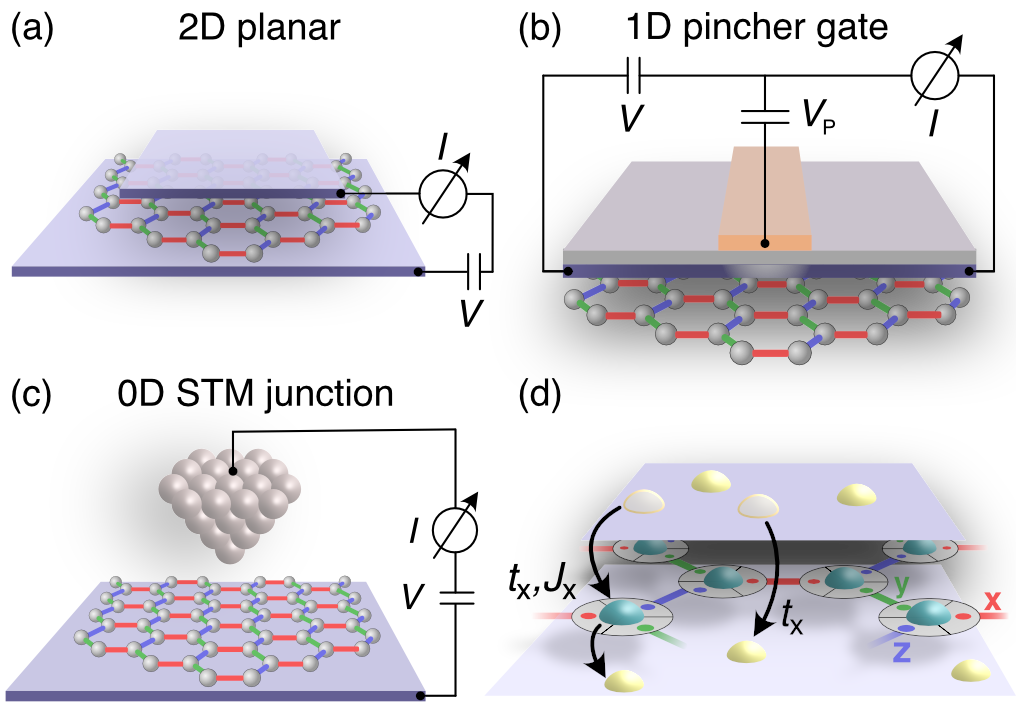}
    \caption{(a) 2D planar tunnel device geometry; a single or few layer QSL material is sandwiched between two metallic 2D electrodes, e.g. graphene \cite{geim2018magnon,carrega2020tunneling}. The tunnel current, $I$, measured as a function of the applied bias voltage, $V$ provides insight on the contributing tunnel processes. (b) 1D pincher gate geometry; a lateral 1D tunnel junction geometry can be created on top of a QSL bulk crystal. A pincher gate can induce an electrically insulating region in a suitable electrode material, e.g.~bilayer graphene \cite{geim2007BLG} with a displacement field, by applying a voltage, $V_{\rm P}$, to serve as 1D tunnel barrier. (c) Zero-dimensional tunnel junction between an STM tip and an electrically conducting substrate, which supports a monolayer QSL material on its surface. (d) Illustration of the relevant tunnel processes in a M-QSL-M geometry. The electron can tunnel either elastically with amplitude $t_{\rm x}$ or inelastically with amplitude $J_{\rm x}$, undergoing spin-flip scattering at the fractionalized spin-degree of freedom of the QSL.}
    \label{fig:1}
\end{figure}

On the experimental side, evidence for these emergent quasiparticles is generally rare \cite{Singh2012relevance, Balz2016physical, Cheng2011high, Paddison20176cont}. The search for material realizations of the Kitaev model has focused on Mott-Hubbard systems with partially filled $t_{2\rm g}$ levels and strong spin-orbit coupling  \cite{jackeli2009mott}. Examples encompass the iridates, such as $\alpha$-Na$_2$IrO$_3$ and $\alpha$-Li$_2$IrO$_3$, with effective spin $1/2$ moments on a honeycomb lattice and bond directional Kitaev interactions. While many of these compounds were found to exhibit long-range magnetic order \cite{chun2015direct}, hydrogen intercalation appears to stabilize the QSL state~\cite{KitagawaTakagi2018}. The layered transition-metal trihalide $\alpha$-RuCl$_3$ \cite{Plumb2014}, with similar properties to those of the iridate honeycomb materials, has been gaining traction in the community as a candidate Kitaev QSL material. Most prominently, recent results from neutron scattering experiments on this compound suggest a magnetically disordered state \cite{banerjee2016proximate, banerjee2017neutron}, consistent with the observation of a half-integer thermal quantum Hall effect at finite magnetic fields~\cite{kasahara2018majorana}--a telltale sign of a chiral Majorana boundary mode~\cite{Kitaev2006}. Nevertheless, the charge-less character of these emergent quasiparticles and the electrically insulating bulk of materials in the QSL state overall limit the range of suitable measurement techniques and, in particular, render their detection in electrical transport measurements challenging \cite{Aasen2020eletrical}.

Previous electron tunneling experiments in planar tunnel junctions made from exfoliated 2D materials established a new means to investigate the magnetic properties of atomically thin insulating materials, 
by using them as tunnel barriers between two electrically conducting graphite electrodes \cite{geim2018magnon, Klein2018probing}. Leveraging the electrically insulating behavior of QSL materials for their application as a tunnel barrier, this concept can be naturally extended to the investigation of their charge-neutral quasiparticle excitations. The tunneling electron can undergo inelastic spin scattering at the fractionalized spin states of the QSL, potentially leaving distinct spectroscopic fingerprints in the electron tunnel characteristics, while its charge degree of freedom only participates in the creation of eletron-hole pairs in the tunnel junction electrodes. From a practical perspective, the Kitaev QSL candidate material $\alpha$-RuCl$_3$ can be exfoliated into the monolayer limit \cite{zhou2019possible, mashhadi2019spin}, and it, therefore, offers direct avenues to explore inelastic spin scattering at the Majorana spinon mode in similar planar device structures \cite{carrega2020tunneling}. However, little is known about influence of the tunnel junction geometry and the electronic properties of the metallic leads on this inelastic spin scattering, and most importantly, under which circumstances this process produces a signal strong enough to be detected in an experiment.

In this letter, we methodologically investigate the general spectroscopic tunneling characteristics of M-QSL-M tunnel junctions, which are formed between two metallic electrodes (M) separated by a thin, electrically insulating QSL barrier, in different experimentally relevant geometries (Fig.\,\ref{fig:1}(a)-(c)). For our theoretical analysis, we consider a single layer Kitaev QSL as the tunnel barrier, and we develop the full DC and AC bias voltage-dependent tunnel conductance expressions, including both scalar and spin-flip contributions, as a function of the QSL spin structure factors \cite{KnolleMoessner2014}. For all investigated junction geometries, we find that spin-flip scattering at gapless Majorana spinon modes yields unique features in the DC and AC electron tunneling spectra, by which this process, as we will show, can be distinguished from scattering at magnons modes in magnetically ordered media. Our analyses further reveal that this effect will appear most prominently as a spectral gap at small bias voltages in the DC tunnel characteristics of planar tunnel junction geometries using 2D metallic leads, e.g. those made from graphene \cite{geim2018magnon}, where scalar contributions to the electric tunnel conductance vanish. We further discuss the effects of quantum Hall states in the 2D electrodes on the tunnel spectra, potentially arising from the sizable magnetic field required to drive $\alpha$-RuCl$_3$ into the QSL phase \cite{kasahara2018majorana}, for the various device geometries. We emphasize that the results on the spectral characteristics of M-QSL-M tunnel junctions, derived in this work for the case of the Kitaev QSL, can be generalized to other
QSL materials.

\textit{\textcolor{blue}{Model.}} A simple physical model to describe the M-QSL-M junctions presented in Fig.~\ref{fig:1} is given by $H = H_{\rm leads} + H_{\rm tun}+H_{\rm QSL}$, where
\begin{subequations}
\begin{eqnarray}
H_{\rm leads} &=& \sum_{\xi = 1, 2} \sum_\sigma \int \frac{d^2k}{(2\pi)^2} c_{{\bf k},\sigma, \xi}^\dagger [\epsilon({\bf k}) - \mu]c_{\bf k,\sigma, \xi},\\
H_{\rm tun} &=& \sum_{\v x} \sum_{\sigma, \sigma'} \left[ t_{\v x} \delta_{\sigma \sigma'} + J_{\v x} \vec \sigma_{\sigma, \sigma'} \cdot \hat{\vec S}({\bf x}) \right ] \notag \\
&& {\left [c^\dagger_{{\bf x} 1 \sigma} c_{{\bf x} 2 \sigma'}e^{i eV t} + H.c. \right ]}.
\end{eqnarray}
\label{eq:H0}
$H_{\rm leads}$ describes the tunnel junction leads, $H_{\rm tun}$ the tunnel process and $H_{\rm QSL}$ the QSL serving as the tunnel barrier. The index $\xi = 1,2$ labels the leads, $\v x$ runs over the lattice sites of the quantum magnet, $\hat{\vec{S}}(\v x)$ is the spin-operator of the QSL
at site $\v x$ and $\vec \sigma$ denotes the spin of the tunneling electron. The first (second) term $t_{\v x}$ ($J_{\v x}$) in the tunneling matrix elements stems from electrons passing through the QSL without affecting the spin configuration (while creating a spin-flip) at site ${\v x}$ (Fig.\,\ref{fig:1}(d)). When the QSL material is placed on a metallic substrate, one may also expect Kondo-like spin-spin interactions with the electron in the underlying metals. Here we consider the case when these are irrelevant, and we, therefore, do not include such interactions. This assumption is well justified when the magnetic correlations are stronger than the Kondo interaction; for example, in Kitaev materials, the vison gap prevents a weak coupling Kondo effect~\cite{SeifertVojta2018,ChoiKim2018}.

Here we study three experimental setups: First, we study planar 2D to 2D tunneling $t_{\v x} = t_0 , J_{\v x} = J_0$ across the M-QSL-M junction, Fig.\,\ref{fig:1}(a). Second, we study a one-dimensional tunneling constriction, $t_{\v x} = t_0 \delta_{x,0} , J_{\v x} = J_0 \delta_{x,0}$, Fig.\,\ref{fig:1}(b). Without magnetic field this setup corresponds to lateral tunneling between two-dimensional electron gases (2DEGs), but in the presence of a sufficiently strong field it represents tunneling between quantum Hall edge states. Third, we consider a zero-dimensional point contact, $t_{\v x} = t_0 \delta_{\v x,0}, J_{\v x} = J_0 \delta_{\v x,0}$, Fig.\,\ref{fig:1}(c), which describes the physics of a classic scanning tunneling microscope (STM) experiment.

As the physical process of tunneling through a Mott-insulating material involves the virtual double occupancy of sites, the microscopic expressions are $J_{\v x} \sim t_{\v x} \sim V^2/U$, where $V$ is the hybridization between the conduction electrons of the leads and the localized electrons in the QSL, and $U$ is the Mott-Hubbard gap of the latter. Particularly in the case of the 1D tunneling barrier, Fig.\,\ref{fig:1}(b), $t_{\v x}$  may acquire an additional contribution from direct tunneling between the leads, in which case $\vert t_{\v x}\vert  \gg \vert J_{\v x} \vert$.

For the analytical study of tunneling through a QSL state, we specifically focus on the case of the exactly solvable Kitaev model~\cite{Kitaev2006}; in this case
\begin{equation}
    H_{\rm QSL}= \sum_{i = x,y,z} K_i \sum_{\langle {\bf x},{\bf x}'\rangle_{i}} \hat S_{i}({\bf x}) \hat S_{i}({\bf x}'). \label{eq:Kitaev}
\end{equation}
\end{subequations}
The interactions of the Kitaev model are bond-directed Ising interactions, as displayed in Fig.\,\ref{fig:1}(d). In the following, we concentrate on the isotropic limit of the model $K_{x} = K_{y} = K_{z}=K$. 

We conclude this section by listing the assumptions behind our calculations: We consider the limit when the Fermi wave length $\lambda_{\rm F}$ of the metallic leads exceeds the lattice constant $a$ of the magnet such that a continuum treatment of the leads is justified. Except the case of the STM tip electrode in Fig\,\ref{fig:1}(c), it is furthermore important that the leads are strictly two dimensional, as can be practically realized, e.g., by using graphene electrodes \cite{geim2018magnon}. Finally, we disregard Umklapp scattering in the section on planar tunneling, which is a good approximation when the unit cell of the magnet equals or exceeds the unit cell of the materials at the leads (this is the case, e.g., for $\alpha$-RuCl$_3$ and graphene~\cite{winter2016challenges}). 

\textit{\textcolor{blue}{Tunneling current.}}
The leading order tunneling current, $I = I^{\rm el} +  I^{\rm inel}$, is given by contributions from from elastic, $I^{\rm el}$, and inelastic, $I^{\rm inel}$, tunneling processes, respectively. The first contribution to the current reflects the standard tunnel current across the junction,
\begin{subequations}
\begin{align}
I^{\rm el.} &= 2\pi \frac{e}{h}\sum_{\v x, \v x'}  t_{\v x}t_{\v x'} [\mathcal A^{1^\dagger 2}_{\v x, \v x'}(eV) -1 \leftrightarrow  2]. \label{eq:It}
\end{align}
We denote the the spectral weight of a particle-hole pair with particle (hole) in 
electrode 1 (electrode 2) by $\mathcal A^{1^\dagger 2}_{\v x, \v x'}(E)$, where $E=eV$ is the energy, and $V$ is the bias voltage across the junction. The current presented in Eq.~\eqref{eq:It} can be interpreted as a difference of Fermi-golden rule rates, where the matrix element is encoded in the spatial dependence of $t_{\v x}, \mathcal A^{1^\dagger 2}_{\v x, \v x'}(E)$. 

\begin{figure}
    \centering
    \includegraphics[width = .45 \textwidth]{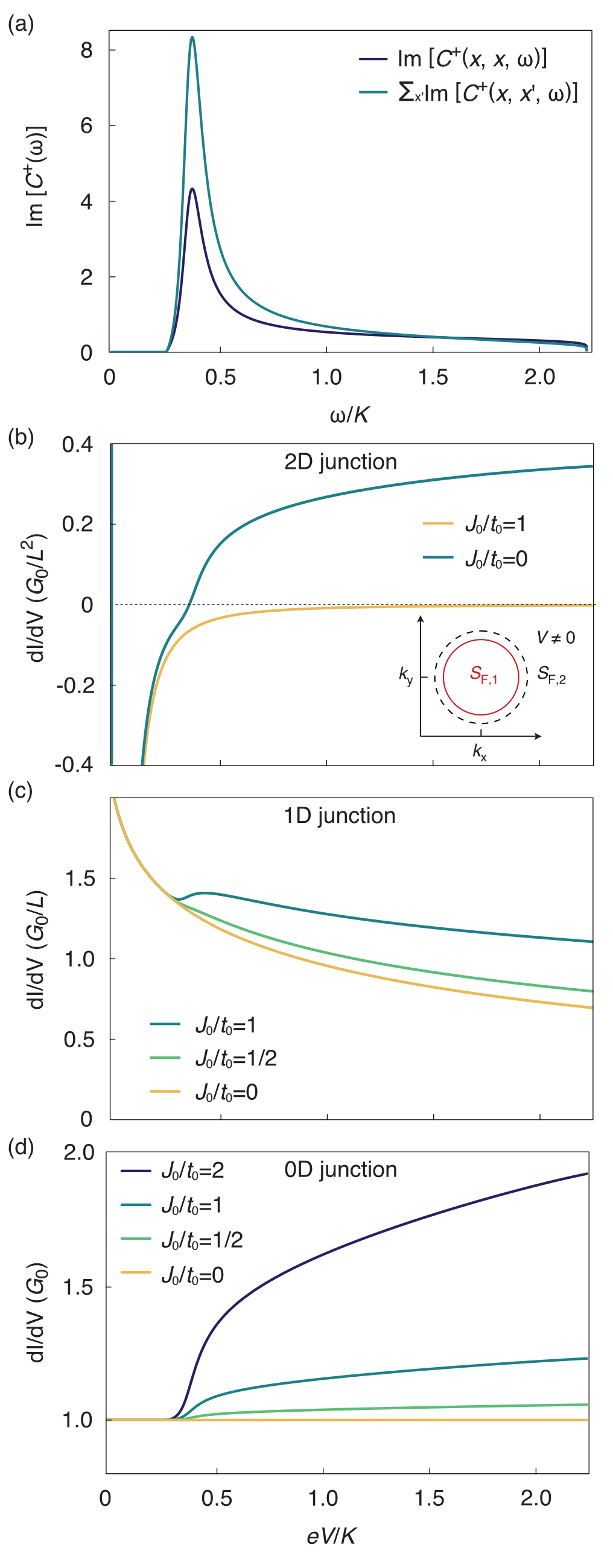}
    \caption{(a) Short-ranged dynamical spin susceptibility, $\Im [C^+(\omega)]$,~\cite{Baskaran2007, KnolleMoessner2014} as a function of frequency, $\omega$, normalized by the Kitaev interaction $K$. (b)-(d) Calculated $dI/dV$-spectra for electron tunneling across junctions of different geometries at different $J_0/t_0$ ratios as a function of the applied bias voltage, $V$. The case of purely elastic electron tunneling corresponds to $J_0/t_0$=0. The inset in (b) depicts the mismatch between the Fermi surfaces, $S_{\rm F, 1}$ and $S_{\rm F, 2}$, between the electrode 1 and 2, respectively at $V\neq0$.}
    \label{fig:2}
    \label{fig:SpectraAndDiDV}
\end{figure}

We now address the second, inelastic contribution to the tunnel current, which can be expressed as~\cite{rossier2009theo, Balatsky2010theo}
\begin{align}
I^{\rm inel.} &= \frac{e}{h} \sum_{\v x, \v x'}\int {d\omega} {J_{\v x} J_{\v x'}} \Big \lbrace  \mathcal A^{\rm spin}_{\v x, \v x'}(eV- \omega) \mathcal A^{1^\dagger 2}_{\v x, \v x'}(\omega) \notag \\
&\times \left [n(eV - \omega) - n(\omega)\right ] - 1 \leftrightarrow  2 \Big \rbrace .
\end{align}
\label{eq:IMainText}
\end{subequations}
As for the elastic current in Eq.~\eqref{eq:It}, the inelastic current is generated by the creation of particle-hole pairs with charges on opposite sides of the junction. In contrast to the scalar contribution, $I^{\rm inel.}$ corresponds to inelastic scattering: The electron-hole pair deposits energy into the spin-system during the tunneling process (both spin-conserving and spin-flip processes are included). This amplitude is weighted by the difference in occupation of the spin and particle-hole modes ($n(E)$ is the Bose-Einstein distribution), and most importantly, by the spectral weight of the spin excitations $\mathcal A^{\rm spin}_{\v x, \v x'}(E) =  -2 \Im C^+(\v x, \v x'; E)$, where
\begin{equation}
 C^+(\v x, \v x'; t,t')= -i \theta(t-t') \sum_{i = x,y,z}  \langle [\hat{S}_i(\v x, t),\hat{S}_i(\v x',t')] \rangle.
\end{equation}
One of the defining characteristics of QSLs is the absence of long-range order, and we focus on cases in which the retarded spin susceptibility, $C^+(\v x, \v x'; t,t') = C^+(\v x- \v x'; t-t')$, is exponentially short ranged both in space and time (i.e.~we disregard algebraic spin liquids~\cite{RantnerWen2001}). When the correlation length is small as compared to the Fermi wave length, the inelastic contribution to the differential tunnel conductance, $dI^{\rm inel}/dV$ at zero temperature can, therefore, be simplified to \cite{SuppMat}
\begin{eqnarray}
\frac{d I^{\rm inel}}{d V} = -{G_0} \sum_{\v x, \v x'} \frac{J_{\v x} J_{\v x'}}{ t_0^2}\int_0^{eV} (d\omega) \Im C^+(\v x- \v x'; E). \label{eq:IJ}
\end{eqnarray}
Here, $G_0\propto t_0^2$ is the dimensionless conductance of a point contact~\cite{SuppMat}. The integral in Eq.~\eqref{eq:IJ} is largely independent of the tunnel junction geometry, and it can be evaluated on the basis of the short ranged spin correlator, $\Im C^+(\v x, \v x'; E)$. In the specific case of the Kitaev spin-liquid, where $\Im C^+(\v x, \v x'; E)$ can be derived analytically~\cite{Baskaran2007,KnolleMoessner2014, SuppMat}, only onsite and nearest neighbour correlators are non-zero~\cite{Baskaran2007}, Fig.~\ref{fig:SpectraAndDiDV}(a) ~\cite{KnolleMoessner2014}. The gap $\sim 0.26 K$ in the spectrum is a manifestation of absent spin order, and it results from creating virtual excitations of the $\mathbb Z_2$ gauge field (`visons'). Beyond this excitation gap, the continuum of Majorana spinons appears as a broad hump. The prefactor to the integral in Eq.~\eqref{eq:IJ} depends on the tunnel-junction geometry, and it acquires a scaling, $\sim L^{d}$ ($L$ denotes the linear dimension of the junction and $d$ the dimension of the tunnel electrode), due to the sum over the mean positions.

We plot the calculated $dI/dV$-spectra in Fig.~\ref{fig:2}, containing both elastic and inelastic contributions, for the different device geometries in Fig.~\ref{fig:1}(a)-(c). These spectra were obtained for $\lambda_{\rm F}/ a= 2\pi, E_{\rm F}/K = 2.5, \Gamma/K =1/1000$, where $E_{\rm F}$ ($\Gamma$) is the Fermi energy (quasiparticle decay rate), and for different values of the tunnel coupling ratio, $t_0/J_0$. In Fig.~\ref{fig:2}, we present results for the DC tunneling experiment only; yet it bears noting that we obtain qualitatively similar characteristics for the AC tunneling conductance, $\Re [G(\Omega)]$~\cite{SuppMat}, whose properties could be probed using Terahertz techniques \cite{wang2017magnetic, little2017antiferromagnetic}

\textit{\textcolor{blue}{Discussion of M-QSL-M setups.}} 
The comparison of the calculated $dI/dV$-spectra in Fig.\,\ref{fig:2}(b)-(d) shows that the inelastic scattering of tunneling electrons off
the itinerant spinon mode of the QSL yields a finite contribution to the tunnel conductance for all junction geometries (for $J_{\v x}/t_{\v x}\neq0$). A closer inspection, however, reveals that the relative contribution of this inelastic channel to the total tunnel conductance varies significantly between the respective cases, as we will discuss in the following.

The planar 2D tunnel junction geometry~\cite{carrega2020tunneling}, with metallic 2DEGs in the junction electrodes, Fig.\,\ref{fig:1}(b), appears particularly well suited for the investigation of the
spin-flip tunneling process. At $J_{0}/t_{0}\neq0$, this inelastic channel results in a prominent bump with an onset at finite voltage in the $dI/dV$ spectrum. By contrast, the elastic channel (cf. curve at $J_{0}/t_{0}=0$) remains largely suppressed at $V\neq0$. Owing to the circular Fermi surfaces, $S_{\rm F, 1}$ and $S_{\rm F, 2}$, of the 2DEGs in the top ($\xi = 1$) and bottom ($\xi = 2$) 
electrodes, respectively, momentum and energy conservation cannot be fulfilled simultaneously at $V\neq0$ (cf. inset Fig.\,\ref{fig:2}(b)). Therefore, the elastic channel only contributes via a Lorentzian current peak centered at $V=0$, the width of which is determined by the quasiparticle lifetime, resulting in a negative contribution to the $dI/dV$-spectrum at finite bias voltages~\cite{MurphyWest1995}.

\begin{figure}
    \centering
  \includegraphics[width = .45\textwidth]{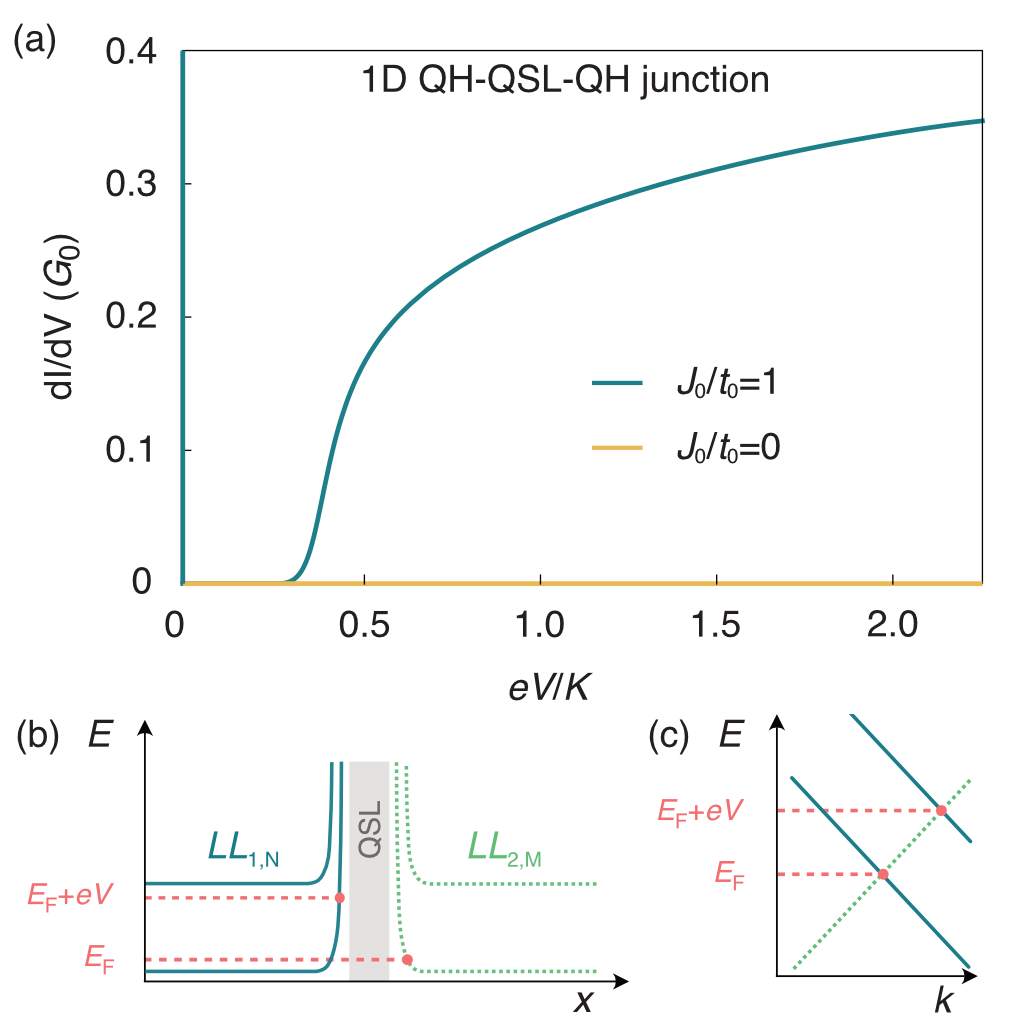}
    \caption{(a) Calculated $dI/dV$-spectrum for electron tunneling between 1D chiral quantum Hall edge states across a 1D tunnel junction for different $J_0/t_0$ ratios as a function of the applied bias voltage, $V$. The case of purely elastic electron tunneling corresponds to $J_0/t_0=0$. (b) Real space energy diagram in the quantum Hall regime (denoted $LL_{1,\rm N}$ and $LL_{2,\rm M}$, where $N,M$ are Landau level indices) of the two tunneling electrodes 1 and 2, respectively. (c) Momentum space diagram of the chiral edge states of $LL_{1,{\rm N}=1}$ and $LL_{2,{\rm M}=1}$ with and without an applied bias $V$.}
    \label{fig:3}
\end{figure}

The lateral 1D tunnel junction geometry, Fig.\,\ref{fig:1}(b), shows fundamentally different $dI/dV$-spectrum characteristics, Fig.\,\ref{fig:2}(c). Such a device could be possibly realized with a 1D pincher gate on top of a 2D semiconductor to electrostatically define a 1D insulating region, serving as the tunnel barrier. A benefit of this geometry is the possibility to place the tunnel electrodes directly on the surface of bulk crystals, which likely expands the range of material candidates as it circumvents challenges related to monolayer exfoliation and unwanted doping~\cite{BiswasValenti2018,mashhadi2019spin}. However, the dominant logarithmic contribution to the $dI/dV$-spectrum originates from elastic tunneling between the 2DEGs, whereas the contribution from spin-flip tunneling is comparably small, rendering its detection presumably challenging.

The third geometry, a 0D tunnel junction, can be formed between an atomically sharp tip of a scanning tunneling microscope and a 2D metallic substrate, which supports the thin QSL material layer, Fig.\,\ref{fig:1}(c). Atomic-scale resolution combined with the ability to distinguish spectral features of the surface from the edge has inspired recent proposals to study QSL spinon modes and chiral Majorana edge modes in such STM setups \cite{chen2020impurity, feldmeier2020local}. Nevertheless, our theoretical analysis of this experimental geometry reveals that a constant background in the $dI/dV$-spectrum, which results from elastic tunneling into the metallic substrate, could render the detection of spin-flip scattering at the Majorana spinon modes challenging over a wide parameter range, Fig.\,\ref{fig:2}(d). Only for $J_{\v x}/t_{\v x}>1$ a significant bump in the $dI/dV$-spectrum develops. On the other hand, the continuous tunability of the STM tip-sample distance could serve as valuable tuning knob to test the evolution of this spectral feature as a function of the STM tunnel junction transparency. Hence, inelastic tunneling with an STM could present an attractive alternative to the studies of QSL states using non-local transport geometries, Fig.\,\ref{fig:1}(a) and (b), not least in view of the recent advances in epitaxial growth of the non-Kitaev QSL candidate materials 1T-TaS$_2$ and 1T-TaSe$_2$ \cite{law20171t, KratochvilovaPark2017, nakata2018selective, lin2018growth, lin2020scanning, chen2020mott}.

\textit{\textcolor{blue}{Quantum Hall regime.}} Up to now, we considered experimental scenarios, in which the electrodes, except for the case of a 0D STM geometry, can be described by a metallic
2DEG. This picture does not, however, always hold. In particular in the case of $\alpha$-RuCl$_3$, the strong out-of-plane magnetic field required to engender the putative QSL state \cite{kasahara2018majorana} induces Landau quantization in the 2D electrodes. Our analysis shows that the presence of Landau levels will have a profound influence on the spectral tunnel characteristics for the 2D and 1D tunnel junction geometry, cf.~Fig.\,\ref{fig:1}(a) and (b).

Previous experiments on 2D planar tunnel junctions show that the presence of quantum Hall states in the 2D graphene electrodes results in a complex $dI/dV$-spectrum \cite{geim2018magnon}. It arises due to the discrete Landau-level spectra, and it presumably renders the observation of tunneling signatures of Majorana spinons \cite{carrega2020tunneling} 
challenging.

By contrast, we establish the case of the 1D lateral tunnel junction in the presence of quantum Hall edge states as a setup, which favors the detection of inelastic spin-flip scattering in the $dI/dV$-spectrum, Fig.\,\ref{fig:3}(a). In the limit of $K/\omega_{\rm c}\ll1$, when the cyclotron frequency, $\omega_{\rm c}$, exceeds the Kitaev coupling, tunneling between the chiral edge modes in both junction electrodes, Fig.\,\ref{fig:3}(b) and (c), results in a constant tunnel current, $I^{\rm el.} \propto G_0 L \text{sign}(eV)$
\cite{BoeseZuelicke2001,SuppMat}. 
Hence, in the case of a 1D QH-QSL-QH tunnel junction, the entire spectral weight in the $dI/dV$-spectrum, Fig.\,\ref{fig:3}(a) at $V\neq0$, arises from inelastic scattering off the spinon modes, providing a strong experimental signature.

\begin{figure}
    \centering
  \includegraphics[width=.45\textwidth]{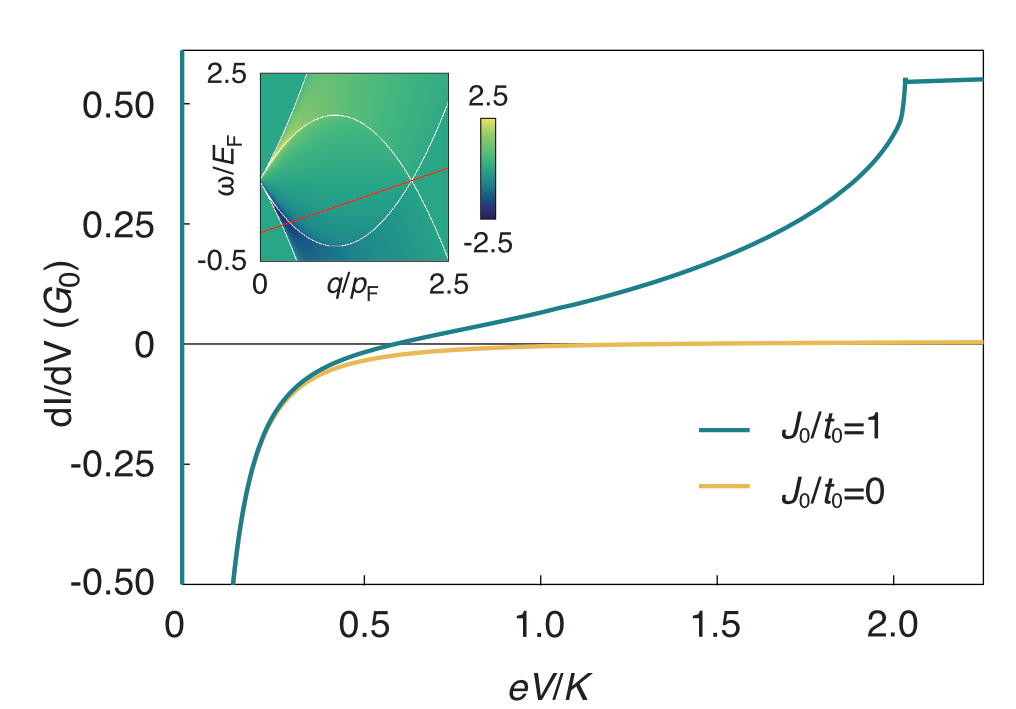}
    \caption{Tunneling spectrum of a planar metal-N{\'e}el antiferromagnet-metal junction. The power-law contribution of inelastic spin-scattering is clearly distinct from Fig.~\ref{fig:2}(b) and substantially less pronounced than the QSL signature. Here, $K = v_s/a$, where $v_s$ is the magnon velocity and all other parameters are as in Fig.~\ref{fig:2}. When $eV/K = 4\pi a/\lambda_F =2$, the magnon dispersion intersects a non-analyticity (inset) in the particle-hole continuum, which manifests itself as a kink.}
    \label{fig:AFM}
\end{figure}

\textit{\textcolor{blue}{Conclusion.}} In this letter, we have presented an extensive comparative study of tunneling signatures for various metal$-$quantum spin-liquid$-$metal junction geometries. The distinguishing feature of tunneling across a QSL is the observation of a spectral gap and a subsequent bump in the $dI/dV$-spectrum at small bias voltages, which is induced by inelastic spin scattering off the fractionalized spinon mode. These  features are most strongly pronounced both for tunneling across planar 2D junctions, Fig.~\ref{fig:2}(b), and for tunneling across a lateral 1D junction, with the electrodes in the quantum Hall regime, Fig.~\ref{fig:3}(a). Albeit we considered the exactly solvable Kitaev model for the QSL \cite{Kitaev2006}, the main results of our analysis can be generalized to other QSL materials with short range spin correlations, too.

We differentiate this behavior from the tunneling signatures of phonons and magnons, and we conclude with a direct comparison to a 2D planar metal-antiferromagnet-metal junction. We concentrate on fluctuations about a N\'eel state on a hexagonal lattice~\cite{SuppMat}, for which the linear magnon spectrum leads to
a cubic inelastic tunneling current $I^{\rm inel.} \sim V^3$.
Contrary to the QSL case, this results in a smooth, quadratic contribution to the $dI/dV$-spectrum at small bias voltages in the absence of an applied magnetic field, Fig.~\ref{fig:AFM}. 

\textit{\textcolor{blue}{Note.}} During the preparation of our manuscript we became aware of related work that focuses on the specific case of a 0D STM junction \cite{feldmeier2020local}.

\textit{\textcolor{blue}{Acknowledgments.}} It is a pleasure to thank P.~P.~Orth for useful discussions. E.~J.~K. acknowledges support by DOE Basic Energy Sciences grant DE-FG02-
99ER45790. M.~T.~R. acknowledges support from the MIT Pappalardo Fellowship. B.~J. acknowledges funding from the Alexander-von-Humboldt foundation through a postdoctoral fellowship.

\bibliography{Bibliography}

\begin{thebibliography}{49}%
\makeatletter
\providecommand \@ifxundefined [1]{%
 \@ifx{#1\undefined}
}%
\providecommand \@ifnum [1]{%
 \ifnum #1\expandafter \@firstoftwo
 \else \expandafter \@secondoftwo
 \fi
}%
\providecommand \@ifx [1]{%
 \ifx #1\expandafter \@firstoftwo
 \else \expandafter \@secondoftwo
 \fi
}%
\providecommand \natexlab [1]{#1}%
\providecommand \enquote  [1]{``#1''}%
\providecommand \bibnamefont  [1]{#1}%
\providecommand \bibfnamefont [1]{#1}%
\providecommand \citenamefont [1]{#1}%
\providecommand \href@noop [0]{\@secondoftwo}%
\providecommand \href [0]{\begingroup \@sanitize@url \@href}%
\providecommand \@href[1]{\@@startlink{#1}\@@href}%
\providecommand \@@href[1]{\endgroup#1\@@endlink}%
\providecommand \@sanitize@url [0]{\catcode `\\12\catcode `\$12\catcode
  `\&12\catcode `\#12\catcode `\^12\catcode `\_12\catcode `\%12\relax}%
\providecommand \@@startlink[1]{}%
\providecommand \@@endlink[0]{}%
\providecommand \url  [0]{\begingroup\@sanitize@url \@url }%
\providecommand \@url [1]{\endgroup\@href {#1}{\urlprefix }}%
\providecommand \urlprefix  [0]{URL }%
\providecommand \Eprint [0]{\href }%
\providecommand \doibase [0]{http://dx.doi.org/}%
\providecommand \selectlanguage [0]{\@gobble}%
\providecommand \bibinfo  [0]{\@secondoftwo}%
\providecommand \bibfield  [0]{\@secondoftwo}%
\providecommand \translation [1]{[#1]}%
\providecommand \BibitemOpen [0]{}%
\providecommand \bibitemStop [0]{}%
\providecommand \bibitemNoStop [0]{.\EOS\space}%
\providecommand \EOS [0]{\spacefactor3000\relax}%
\providecommand \BibitemShut  [1]{\csname bibitem#1\endcsname}%
\let\auto@bib@innerbib\@empty
\bibitem [{\citenamefont {Anderson}(1973)}]{anderson1973resonating}%
  \BibitemOpen
  \bibfield  {author} {\bibinfo {author} {\bibfnamefont {P.~W.}\ \bibnamefont
  {Anderson}},\ }\href
  {https://www.sciencedirect.com/science/article/pii/0025540873901670}
  {\bibfield  {journal} {\bibinfo  {journal} {Materials Research Bulletin}\
  }\textbf {\bibinfo {volume} {8}},\ \bibinfo {pages} {153} (\bibinfo {year}
  {1973})}\BibitemShut {NoStop}%
\bibitem [{\citenamefont {Kitaev}(2006)}]{Kitaev2006}%
  \BibitemOpen
  \bibfield  {author} {\bibinfo {author} {\bibfnamefont {A.}~\bibnamefont
  {Kitaev}},\ }\href
  {https://www.sciencedirect.com/science/article/abs/pii/S0003491605002381}
  {\bibfield  {journal} {\bibinfo  {journal} {Annals of Physics}\ }\textbf
  {\bibinfo {volume} {321}},\ \bibinfo {pages} {2} (\bibinfo {year}
  {2006})}\BibitemShut {NoStop}%
\bibitem [{\citenamefont {Nayak}\ \emph {et~al.}(2008)\citenamefont {Nayak},
  \citenamefont {Simon}, \citenamefont {Stern}, \citenamefont {Freedman},\ and\
  \citenamefont {Sarma}}]{nayak2008non}%
  \BibitemOpen
  \bibfield  {author} {\bibinfo {author} {\bibfnamefont {C.}~\bibnamefont
  {Nayak}}, \bibinfo {author} {\bibfnamefont {S.~H.}\ \bibnamefont {Simon}},
  \bibinfo {author} {\bibfnamefont {A.}~\bibnamefont {Stern}}, \bibinfo
  {author} {\bibfnamefont {M.}~\bibnamefont {Freedman}}, \ and\ \bibinfo
  {author} {\bibfnamefont {S.~D.}\ \bibnamefont {Sarma}},\ }\href
  {https://journals.aps.org/rmp/abstract/10.1103/RevModPhys.80.1083} {\bibfield
   {journal} {\bibinfo  {journal} {Reviews of Modern Physics}\ }\textbf
  {\bibinfo {volume} {80}},\ \bibinfo {pages} {1083} (\bibinfo {year}
  {2008})}\BibitemShut {NoStop}%
\bibitem [{\citenamefont {Jackeli}\ and\ \citenamefont
  {Khaliullin}(2009)}]{jackeli2009mott}%
  \BibitemOpen
  \bibfield  {author} {\bibinfo {author} {\bibfnamefont {G.}~\bibnamefont
  {Jackeli}}\ and\ \bibinfo {author} {\bibfnamefont {G.}~\bibnamefont
  {Khaliullin}},\ }\href
  {https://journals.aps.org/prl/abstract/10.1103/PhysRevLett.102.017205}
  {\bibfield  {journal} {\bibinfo  {journal} {Physical review letters}\
  }\textbf {\bibinfo {volume} {102}},\ \bibinfo {pages} {017205} (\bibinfo
  {year} {2009})}\BibitemShut {NoStop}%
\bibitem [{\citenamefont {Savary}\ and\ \citenamefont
  {Balents}(2016)}]{SavaryBalents2016}%
  \BibitemOpen
  \bibfield  {author} {\bibinfo {author} {\bibfnamefont {L.}~\bibnamefont
  {Savary}}\ and\ \bibinfo {author} {\bibfnamefont {L.}~\bibnamefont
  {Balents}},\ }\href
  {https://iopscience.iop.org/article/10.1088/0034-4885/80/1/016502} {\bibfield
   {journal} {\bibinfo  {journal} {Reports on Progress in Physics}\ }\textbf
  {\bibinfo {volume} {80}},\ \bibinfo {pages} {016502} (\bibinfo {year}
  {2016})}\BibitemShut {NoStop}%
\bibitem [{\citenamefont {Zhou}\ \emph {et~al.}(2017)\citenamefont {Zhou},
  \citenamefont {Kanoda},\ and\ \citenamefont {Ng}}]{zhou2017quantum}%
  \BibitemOpen
  \bibfield  {author} {\bibinfo {author} {\bibfnamefont {Y.}~\bibnamefont
  {Zhou}}, \bibinfo {author} {\bibfnamefont {K.}~\bibnamefont {Kanoda}}, \ and\
  \bibinfo {author} {\bibfnamefont {T.-K.}\ \bibnamefont {Ng}},\ }\href
  {https://journals.aps.org/rmp/abstract/10.1103/RevModPhys.89.025003}
  {\bibfield  {journal} {\bibinfo  {journal} {Reviews of Modern Physics}\
  }\textbf {\bibinfo {volume} {89}},\ \bibinfo {pages} {025003} (\bibinfo
  {year} {2017})}\BibitemShut {NoStop}%
\bibitem [{\citenamefont {Knolle}\ and\ \citenamefont
  {Moessner}(2019)}]{knolle2019field}%
  \BibitemOpen
  \bibfield  {author} {\bibinfo {author} {\bibfnamefont {J.}~\bibnamefont
  {Knolle}}\ and\ \bibinfo {author} {\bibfnamefont {R.}~\bibnamefont
  {Moessner}},\ }\href
  {https://www.annualreviews.org/doi/abs/10.1146/annurev-conmatphys-031218-013401}
  {\bibfield  {journal} {\bibinfo  {journal} {Annual Review of Condensed Matter
  Physics}\ }\textbf {\bibinfo {volume} {10}},\ \bibinfo {pages} {451}
  (\bibinfo {year} {2019})}\BibitemShut {NoStop}%
\bibitem [{\citenamefont {Ghazaryan}\ \emph {et~al.}(2018)\citenamefont
  {Ghazaryan}, \citenamefont {Greenaway}, \citenamefont {Wang}, \citenamefont
  {Guarochico-Moreira}, \citenamefont {Vera-Marun}, \citenamefont {Yin},
  \citenamefont {Liao}, \citenamefont {Morozov}, \citenamefont {Kristanovski},
  \citenamefont {Lichtenstein}, \citenamefont {Katsnelson}, \citenamefont
  {Withers}, \citenamefont {Mishchenko}, \citenamefont {Eaves}, \citenamefont
  {Geim}, \citenamefont {Novoselov},\ and\ \citenamefont
  {Misra}}]{geim2018magnon}%
  \BibitemOpen
  \bibfield  {author} {\bibinfo {author} {\bibfnamefont {D.}~\bibnamefont
  {Ghazaryan}}, \bibinfo {author} {\bibfnamefont {M.~T.}\ \bibnamefont
  {Greenaway}}, \bibinfo {author} {\bibfnamefont {Z.}~\bibnamefont {Wang}},
  \bibinfo {author} {\bibfnamefont {V.~H.}\ \bibnamefont {Guarochico-Moreira}},
  \bibinfo {author} {\bibfnamefont {I.~J.}\ \bibnamefont {Vera-Marun}},
  \bibinfo {author} {\bibfnamefont {J.}~\bibnamefont {Yin}}, \bibinfo {author}
  {\bibfnamefont {Y.}~\bibnamefont {Liao}}, \bibinfo {author} {\bibfnamefont
  {S.~V.}\ \bibnamefont {Morozov}}, \bibinfo {author} {\bibfnamefont
  {O.}~\bibnamefont {Kristanovski}}, \bibinfo {author} {\bibfnamefont {A.~I.}\
  \bibnamefont {Lichtenstein}}, \bibinfo {author} {\bibfnamefont {M.~I.}\
  \bibnamefont {Katsnelson}}, \bibinfo {author} {\bibfnamefont
  {F.}~\bibnamefont {Withers}}, \bibinfo {author} {\bibfnamefont
  {A.}~\bibnamefont {Mishchenko}}, \bibinfo {author} {\bibfnamefont
  {L.}~\bibnamefont {Eaves}}, \bibinfo {author} {\bibfnamefont {A.~K.}\
  \bibnamefont {Geim}}, \bibinfo {author} {\bibfnamefont {K.~S.}\ \bibnamefont
  {Novoselov}}, \ and\ \bibinfo {author} {\bibfnamefont {A.}~\bibnamefont
  {Misra}},\ }\href {\doibase 10.1038/s41928-018-0087-z} {\bibfield  {journal}
  {\bibinfo  {journal} {Nature Electronics}\ }\textbf {\bibinfo {volume} {1}},\
  \bibinfo {pages} {344} (\bibinfo {year} {2018})}\BibitemShut {NoStop}%
\bibitem [{\citenamefont {Carrega}\ \emph {et~al.}(2020)\citenamefont
  {Carrega}, \citenamefont {Vera-Marun},\ and\ \citenamefont
  {Principi}}]{carrega2020tunneling}%
  \BibitemOpen
  \bibfield  {author} {\bibinfo {author} {\bibfnamefont {M.}~\bibnamefont
  {Carrega}}, \bibinfo {author} {\bibfnamefont {I.~J.}\ \bibnamefont
  {Vera-Marun}}, \ and\ \bibinfo {author} {\bibfnamefont {A.}~\bibnamefont
  {Principi}},\ }\href {https://arxiv.org/abs/2004.13036} {\bibfield  {journal}
  {\bibinfo  {journal} {arXiv preprint arXiv:2004.13036}\ } (\bibinfo {year}
  {2020})}\BibitemShut {NoStop}%
\bibitem [{\citenamefont {Castro}\ \emph {et~al.}(2007)\citenamefont {Castro},
  \citenamefont {Novoselov}, \citenamefont {Morozov}, \citenamefont {Peres},
  \citenamefont {dos Santos}, \citenamefont {Nilsson}, \citenamefont {Guinea},
  \citenamefont {Geim},\ and\ \citenamefont {Neto}}]{geim2007BLG}%
  \BibitemOpen
  \bibfield  {author} {\bibinfo {author} {\bibfnamefont {E.~V.}\ \bibnamefont
  {Castro}}, \bibinfo {author} {\bibfnamefont {K.~S.}\ \bibnamefont
  {Novoselov}}, \bibinfo {author} {\bibfnamefont {S.~V.}\ \bibnamefont
  {Morozov}}, \bibinfo {author} {\bibfnamefont {N.~M.~R.}\ \bibnamefont
  {Peres}}, \bibinfo {author} {\bibfnamefont {J.~M. B.~L.}\ \bibnamefont {dos
  Santos}}, \bibinfo {author} {\bibfnamefont {J.}~\bibnamefont {Nilsson}},
  \bibinfo {author} {\bibfnamefont {F.}~\bibnamefont {Guinea}}, \bibinfo
  {author} {\bibfnamefont {A.~K.}\ \bibnamefont {Geim}}, \ and\ \bibinfo
  {author} {\bibfnamefont {A.~H.~C.}\ \bibnamefont {Neto}},\ }\href {\doibase
  10.1103/PhysRevLett.99.216802} {\bibfield  {journal} {\bibinfo  {journal}
  {Phys. Rev. Lett.}\ }\textbf {\bibinfo {volume} {99}},\ \bibinfo {pages}
  {216802} (\bibinfo {year} {2007})}\BibitemShut {NoStop}%
\bibitem [{\citenamefont {Singh}\ \emph {et~al.}(2012)\citenamefont {Singh},
  \citenamefont {Manni}, \citenamefont {Reuther}, \citenamefont {Berlijn},
  \citenamefont {Thomale}, \citenamefont {Ku}, \citenamefont {Trebst},\ and\
  \citenamefont {Gegenwart}}]{Singh2012relevance}%
  \BibitemOpen
  \bibfield  {author} {\bibinfo {author} {\bibfnamefont {Y.}~\bibnamefont
  {Singh}}, \bibinfo {author} {\bibfnamefont {S.}~\bibnamefont {Manni}},
  \bibinfo {author} {\bibfnamefont {J.}~\bibnamefont {Reuther}}, \bibinfo
  {author} {\bibfnamefont {T.}~\bibnamefont {Berlijn}}, \bibinfo {author}
  {\bibfnamefont {R.}~\bibnamefont {Thomale}}, \bibinfo {author} {\bibfnamefont
  {W.}~\bibnamefont {Ku}}, \bibinfo {author} {\bibfnamefont {S.}~\bibnamefont
  {Trebst}}, \ and\ \bibinfo {author} {\bibfnamefont {P.}~\bibnamefont
  {Gegenwart}},\ }\href {\doibase 10.1103/PhysRevLett.108.127203} {\bibfield
  {journal} {\bibinfo  {journal} {Phys. Rev. Lett.}\ }\textbf {\bibinfo
  {volume} {108}},\ \bibinfo {pages} {127203} (\bibinfo {year}
  {2012})}\BibitemShut {NoStop}%
\bibitem [{\citenamefont {Balz}\ \emph {et~al.}(2016)\citenamefont {Balz},
  \citenamefont {Lake}, \citenamefont {Reuther}, \citenamefont {Luetkens},
  \citenamefont {Sch{\"o}nemann}, \citenamefont {Herrmannsd{\"o}rfer},
  \citenamefont {Singh}, \citenamefont {Nazmul~Islam}, \citenamefont {Wheeler},
  \citenamefont {Rodriguez-Rivera}, \citenamefont {Guidi}, \citenamefont
  {Simeoni}, \citenamefont {Baines},\ and\ \citenamefont
  {Ryll}}]{Balz2016physical}%
  \BibitemOpen
  \bibfield  {author} {\bibinfo {author} {\bibfnamefont {C.}~\bibnamefont
  {Balz}}, \bibinfo {author} {\bibfnamefont {B.}~\bibnamefont {Lake}}, \bibinfo
  {author} {\bibfnamefont {J.}~\bibnamefont {Reuther}}, \bibinfo {author}
  {\bibfnamefont {H.}~\bibnamefont {Luetkens}}, \bibinfo {author}
  {\bibfnamefont {R.}~\bibnamefont {Sch{\"o}nemann}}, \bibinfo {author}
  {\bibfnamefont {T.}~\bibnamefont {Herrmannsd{\"o}rfer}}, \bibinfo {author}
  {\bibfnamefont {Y.}~\bibnamefont {Singh}}, \bibinfo {author} {\bibfnamefont
  {A.~T.~M.}\ \bibnamefont {Nazmul~Islam}}, \bibinfo {author} {\bibfnamefont
  {E.~M.}\ \bibnamefont {Wheeler}}, \bibinfo {author} {\bibfnamefont {J.~A.}\
  \bibnamefont {Rodriguez-Rivera}}, \bibinfo {author} {\bibfnamefont
  {T.}~\bibnamefont {Guidi}}, \bibinfo {author} {\bibfnamefont {G.~G.}\
  \bibnamefont {Simeoni}}, \bibinfo {author} {\bibfnamefont {C.}~\bibnamefont
  {Baines}}, \ and\ \bibinfo {author} {\bibfnamefont {H.}~\bibnamefont
  {Ryll}},\ }\href {\doibase 10.1038/nphys3826} {\bibfield  {journal} {\bibinfo
   {journal} {Nature Physics}\ }\textbf {\bibinfo {volume} {12}},\ \bibinfo
  {pages} {942} (\bibinfo {year} {2016})}\BibitemShut {NoStop}%
\bibitem [{\citenamefont {Cheng}\ \emph {et~al.}(2011)\citenamefont {Cheng},
  \citenamefont {Li}, \citenamefont {Balicas}, \citenamefont {Zhou},
  \citenamefont {Goodenough}, \citenamefont {Xu},\ and\ \citenamefont
  {Zhou}}]{Cheng2011high}%
  \BibitemOpen
  \bibfield  {author} {\bibinfo {author} {\bibfnamefont {J.~G.}\ \bibnamefont
  {Cheng}}, \bibinfo {author} {\bibfnamefont {G.}~\bibnamefont {Li}}, \bibinfo
  {author} {\bibfnamefont {L.}~\bibnamefont {Balicas}}, \bibinfo {author}
  {\bibfnamefont {J.~S.}\ \bibnamefont {Zhou}}, \bibinfo {author}
  {\bibfnamefont {J.~B.}\ \bibnamefont {Goodenough}}, \bibinfo {author}
  {\bibfnamefont {C.}~\bibnamefont {Xu}}, \ and\ \bibinfo {author}
  {\bibfnamefont {H.~D.}\ \bibnamefont {Zhou}},\ }\href {\doibase
  10.1103/PhysRevLett.107.197204} {\bibfield  {journal} {\bibinfo  {journal}
  {Phys. Rev. Lett.}\ }\textbf {\bibinfo {volume} {107}},\ \bibinfo {pages}
  {197204} (\bibinfo {year} {2011})}\BibitemShut {NoStop}%
\bibitem [{\citenamefont {Paddison}\ \emph {et~al.}(2017)\citenamefont
  {Paddison}, \citenamefont {Daum}, \citenamefont {Dun}, \citenamefont
  {Ehlers}, \citenamefont {Liu}, \citenamefont {Stone}, \citenamefont {Zhou},\
  and\ \citenamefont {Mourigal}}]{Paddison20176cont}%
  \BibitemOpen
  \bibfield  {author} {\bibinfo {author} {\bibfnamefont {J.~A.~M.}\
  \bibnamefont {Paddison}}, \bibinfo {author} {\bibfnamefont {M.}~\bibnamefont
  {Daum}}, \bibinfo {author} {\bibfnamefont {Z.}~\bibnamefont {Dun}}, \bibinfo
  {author} {\bibfnamefont {G.}~\bibnamefont {Ehlers}}, \bibinfo {author}
  {\bibfnamefont {Y.}~\bibnamefont {Liu}}, \bibinfo {author} {\bibfnamefont
  {M.~B.}\ \bibnamefont {Stone}}, \bibinfo {author} {\bibfnamefont
  {H.}~\bibnamefont {Zhou}}, \ and\ \bibinfo {author} {\bibfnamefont
  {M.}~\bibnamefont {Mourigal}},\ }\href {\doibase 10.1038/nphys3971}
  {\bibfield  {journal} {\bibinfo  {journal} {Nature Physics}\ }\textbf
  {\bibinfo {volume} {13}},\ \bibinfo {pages} {117} (\bibinfo {year}
  {2017})}\BibitemShut {NoStop}%
\bibitem [{\citenamefont {Chun}\ \emph {et~al.}(2015)\citenamefont {Chun},
  \citenamefont {Kim}, \citenamefont {Kim}, \citenamefont {Zheng},
  \citenamefont {Stoumpos}, \citenamefont {Malliakas}, \citenamefont
  {Mitchell}, \citenamefont {Mehlawat}, \citenamefont {Singh}, \citenamefont
  {Choi} \emph {et~al.}}]{chun2015direct}%
  \BibitemOpen
  \bibfield  {author} {\bibinfo {author} {\bibfnamefont {S.~H.}\ \bibnamefont
  {Chun}}, \bibinfo {author} {\bibfnamefont {J.-W.}\ \bibnamefont {Kim}},
  \bibinfo {author} {\bibfnamefont {J.}~\bibnamefont {Kim}}, \bibinfo {author}
  {\bibfnamefont {H.}~\bibnamefont {Zheng}}, \bibinfo {author} {\bibfnamefont
  {C.~C.}\ \bibnamefont {Stoumpos}}, \bibinfo {author} {\bibfnamefont
  {C.}~\bibnamefont {Malliakas}}, \bibinfo {author} {\bibfnamefont
  {J.}~\bibnamefont {Mitchell}}, \bibinfo {author} {\bibfnamefont
  {K.}~\bibnamefont {Mehlawat}}, \bibinfo {author} {\bibfnamefont
  {Y.}~\bibnamefont {Singh}}, \bibinfo {author} {\bibfnamefont
  {Y.}~\bibnamefont {Choi}},  \emph {et~al.},\ }\href
  {https://www.nature.com/articles/nphys3322} {\bibfield  {journal} {\bibinfo
  {journal} {Nature Physics}\ }\textbf {\bibinfo {volume} {11}},\ \bibinfo
  {pages} {462} (\bibinfo {year} {2015})}\BibitemShut {NoStop}%
\bibitem [{\citenamefont {Kitagawa}\ \emph {et~al.}(2018)\citenamefont
  {Kitagawa}, \citenamefont {Takayama}, \citenamefont {Matsumoto},
  \citenamefont {Kato}, \citenamefont {Takano}, \citenamefont {Kishimoto},
  \citenamefont {Bette}, \citenamefont {Dinnebier}, \citenamefont {Jackeli},\
  and\ \citenamefont {Takagi}}]{KitagawaTakagi2018}%
  \BibitemOpen
  \bibfield  {author} {\bibinfo {author} {\bibfnamefont {K.}~\bibnamefont
  {Kitagawa}}, \bibinfo {author} {\bibfnamefont {T.}~\bibnamefont {Takayama}},
  \bibinfo {author} {\bibfnamefont {Y.}~\bibnamefont {Matsumoto}}, \bibinfo
  {author} {\bibfnamefont {A.}~\bibnamefont {Kato}}, \bibinfo {author}
  {\bibfnamefont {R.}~\bibnamefont {Takano}}, \bibinfo {author} {\bibfnamefont
  {Y.}~\bibnamefont {Kishimoto}}, \bibinfo {author} {\bibfnamefont
  {S.}~\bibnamefont {Bette}}, \bibinfo {author} {\bibfnamefont
  {R.}~\bibnamefont {Dinnebier}}, \bibinfo {author} {\bibfnamefont
  {G.}~\bibnamefont {Jackeli}}, \ and\ \bibinfo {author} {\bibfnamefont
  {H.}~\bibnamefont {Takagi}},\ }\href
  {https://www.nature.com/articles/nature25482} {\bibfield  {journal} {\bibinfo
   {journal} {Nature}\ }\textbf {\bibinfo {volume} {554}},\ \bibinfo {pages}
  {341} (\bibinfo {year} {2018})}\BibitemShut {NoStop}%
\bibitem [{\citenamefont {Plumb}\ \emph {et~al.}(2014)\citenamefont {Plumb},
  \citenamefont {Clancy}, \citenamefont {Sandilands}, \citenamefont {Shankar},
  \citenamefont {Hu}, \citenamefont {Burch}, \citenamefont {Kee},\ and\
  \citenamefont {Kim}}]{Plumb2014}%
  \BibitemOpen
  \bibfield  {author} {\bibinfo {author} {\bibfnamefont {K.~W.}\ \bibnamefont
  {Plumb}}, \bibinfo {author} {\bibfnamefont {J.~P.}\ \bibnamefont {Clancy}},
  \bibinfo {author} {\bibfnamefont {L.~J.}\ \bibnamefont {Sandilands}},
  \bibinfo {author} {\bibfnamefont {V.~V.}\ \bibnamefont {Shankar}}, \bibinfo
  {author} {\bibfnamefont {Y.~F.}\ \bibnamefont {Hu}}, \bibinfo {author}
  {\bibfnamefont {K.~S.}\ \bibnamefont {Burch}}, \bibinfo {author}
  {\bibfnamefont {H.-Y.}\ \bibnamefont {Kee}}, \ and\ \bibinfo {author}
  {\bibfnamefont {Y.-J.}\ \bibnamefont {Kim}},\ }\href {\doibase
  10.1103/PhysRevB.90.041112} {\bibfield  {journal} {\bibinfo  {journal} {Phys.
  Rev. B}\ }\textbf {\bibinfo {volume} {90}},\ \bibinfo {pages} {041112}
  (\bibinfo {year} {2014})}\BibitemShut {NoStop}%
\bibitem [{\citenamefont {Banerjee}\ \emph {et~al.}(2016)\citenamefont
  {Banerjee}, \citenamefont {Bridges}, \citenamefont {Yan}, \citenamefont
  {Aczel}, \citenamefont {Li}, \citenamefont {Stone}, \citenamefont {Granroth},
  \citenamefont {Lumsden}, \citenamefont {Yiu}, \citenamefont {Knolle} \emph
  {et~al.}}]{banerjee2016proximate}%
  \BibitemOpen
  \bibfield  {author} {\bibinfo {author} {\bibfnamefont {A.}~\bibnamefont
  {Banerjee}}, \bibinfo {author} {\bibfnamefont {C.}~\bibnamefont {Bridges}},
  \bibinfo {author} {\bibfnamefont {J.-Q.}\ \bibnamefont {Yan}}, \bibinfo
  {author} {\bibfnamefont {A.}~\bibnamefont {Aczel}}, \bibinfo {author}
  {\bibfnamefont {L.}~\bibnamefont {Li}}, \bibinfo {author} {\bibfnamefont
  {M.}~\bibnamefont {Stone}}, \bibinfo {author} {\bibfnamefont
  {G.}~\bibnamefont {Granroth}}, \bibinfo {author} {\bibfnamefont
  {M.}~\bibnamefont {Lumsden}}, \bibinfo {author} {\bibfnamefont
  {Y.}~\bibnamefont {Yiu}}, \bibinfo {author} {\bibfnamefont {J.}~\bibnamefont
  {Knolle}},  \emph {et~al.},\ }\href
  {https://www.nature.com/articles/nmat4604} {\bibfield  {journal} {\bibinfo
  {journal} {Nature materials}\ }\textbf {\bibinfo {volume} {15}},\ \bibinfo
  {pages} {733} (\bibinfo {year} {2016})}\BibitemShut {NoStop}%
\bibitem [{\citenamefont {Banerjee}\ \emph {et~al.}(2017)\citenamefont
  {Banerjee}, \citenamefont {Yan}, \citenamefont {Knolle}, \citenamefont
  {Bridges}, \citenamefont {Stone}, \citenamefont {Lumsden}, \citenamefont
  {Mandrus}, \citenamefont {Tennant}, \citenamefont {Moessner},\ and\
  \citenamefont {Nagler}}]{banerjee2017neutron}%
  \BibitemOpen
  \bibfield  {author} {\bibinfo {author} {\bibfnamefont {A.}~\bibnamefont
  {Banerjee}}, \bibinfo {author} {\bibfnamefont {J.}~\bibnamefont {Yan}},
  \bibinfo {author} {\bibfnamefont {J.}~\bibnamefont {Knolle}}, \bibinfo
  {author} {\bibfnamefont {C.~A.}\ \bibnamefont {Bridges}}, \bibinfo {author}
  {\bibfnamefont {M.~B.}\ \bibnamefont {Stone}}, \bibinfo {author}
  {\bibfnamefont {M.~D.}\ \bibnamefont {Lumsden}}, \bibinfo {author}
  {\bibfnamefont {D.~G.}\ \bibnamefont {Mandrus}}, \bibinfo {author}
  {\bibfnamefont {D.~A.}\ \bibnamefont {Tennant}}, \bibinfo {author}
  {\bibfnamefont {R.}~\bibnamefont {Moessner}}, \ and\ \bibinfo {author}
  {\bibfnamefont {S.~E.}\ \bibnamefont {Nagler}},\ }\href
  {https://science.sciencemag.org/content/356/6342/1055} {\bibfield  {journal}
  {\bibinfo  {journal} {Science}\ }\textbf {\bibinfo {volume} {356}},\ \bibinfo
  {pages} {1055} (\bibinfo {year} {2017})}\BibitemShut {NoStop}%
\bibitem [{\citenamefont {Kasahara}\ \emph {et~al.}(2018)\citenamefont
  {Kasahara}, \citenamefont {Ohnishi}, \citenamefont {Mizukami}, \citenamefont
  {Tanaka}, \citenamefont {Ma}, \citenamefont {Sugii}, \citenamefont {Kurita},
  \citenamefont {Tanaka}, \citenamefont {Nasu}, \citenamefont {Motome} \emph
  {et~al.}}]{kasahara2018majorana}%
  \BibitemOpen
  \bibfield  {author} {\bibinfo {author} {\bibfnamefont {Y.}~\bibnamefont
  {Kasahara}}, \bibinfo {author} {\bibfnamefont {T.}~\bibnamefont {Ohnishi}},
  \bibinfo {author} {\bibfnamefont {Y.}~\bibnamefont {Mizukami}}, \bibinfo
  {author} {\bibfnamefont {O.}~\bibnamefont {Tanaka}}, \bibinfo {author}
  {\bibfnamefont {S.}~\bibnamefont {Ma}}, \bibinfo {author} {\bibfnamefont
  {K.}~\bibnamefont {Sugii}}, \bibinfo {author} {\bibfnamefont
  {N.}~\bibnamefont {Kurita}}, \bibinfo {author} {\bibfnamefont
  {H.}~\bibnamefont {Tanaka}}, \bibinfo {author} {\bibfnamefont
  {J.}~\bibnamefont {Nasu}}, \bibinfo {author} {\bibfnamefont {Y.}~\bibnamefont
  {Motome}},  \emph {et~al.},\ }\href
  {https://www.nature.com/articles/s41586-018-0274-0} {\bibfield  {journal}
  {\bibinfo  {journal} {Nature}\ }\textbf {\bibinfo {volume} {559}},\ \bibinfo
  {pages} {227} (\bibinfo {year} {2018})}\BibitemShut {NoStop}%
\bibitem [{\citenamefont {Aasen}\ \emph {et~al.}(2020)\citenamefont {Aasen},
  \citenamefont {Mong}, \citenamefont {Hunt}, \citenamefont {Mandrus},\ and\
  \citenamefont {Alicea}}]{Aasen2020eletrical}%
  \BibitemOpen
  \bibfield  {author} {\bibinfo {author} {\bibfnamefont {D.}~\bibnamefont
  {Aasen}}, \bibinfo {author} {\bibfnamefont {R.~S.~K.}\ \bibnamefont {Mong}},
  \bibinfo {author} {\bibfnamefont {B.~M.}\ \bibnamefont {Hunt}}, \bibinfo
  {author} {\bibfnamefont {D.}~\bibnamefont {Mandrus}}, \ and\ \bibinfo
  {author} {\bibfnamefont {J.}~\bibnamefont {Alicea}},\ }\href {\doibase
  10.1103/PhysRevX.10.031014} {\bibfield  {journal} {\bibinfo  {journal} {Phys.
  Rev. X}\ }\textbf {\bibinfo {volume} {10}},\ \bibinfo {pages} {031014}
  (\bibinfo {year} {2020})}\BibitemShut {NoStop}%
\bibitem [{\citenamefont {Klein}\ \emph {et~al.}(2018)\citenamefont {Klein},
  \citenamefont {MacNeill}, \citenamefont {Lado}, \citenamefont {Soriano},
  \citenamefont {Navarro-Moratalla}, \citenamefont {Watanabe}, \citenamefont
  {Taniguchi}, \citenamefont {Manni}, \citenamefont {Canfield}, \citenamefont
  {Fern{\'a}ndez-Rossier},\ and\ \citenamefont
  {Jarillo-Herrero}}]{Klein2018probing}%
  \BibitemOpen
  \bibfield  {author} {\bibinfo {author} {\bibfnamefont {D.~R.}\ \bibnamefont
  {Klein}}, \bibinfo {author} {\bibfnamefont {D.}~\bibnamefont {MacNeill}},
  \bibinfo {author} {\bibfnamefont {J.~L.}\ \bibnamefont {Lado}}, \bibinfo
  {author} {\bibfnamefont {D.}~\bibnamefont {Soriano}}, \bibinfo {author}
  {\bibfnamefont {E.}~\bibnamefont {Navarro-Moratalla}}, \bibinfo {author}
  {\bibfnamefont {K.}~\bibnamefont {Watanabe}}, \bibinfo {author}
  {\bibfnamefont {T.}~\bibnamefont {Taniguchi}}, \bibinfo {author}
  {\bibfnamefont {S.}~\bibnamefont {Manni}}, \bibinfo {author} {\bibfnamefont
  {P.}~\bibnamefont {Canfield}}, \bibinfo {author} {\bibfnamefont
  {J.}~\bibnamefont {Fern{\'a}ndez-Rossier}}, \ and\ \bibinfo {author}
  {\bibfnamefont {P.}~\bibnamefont {Jarillo-Herrero}},\ }\href {\doibase
  10.1126/science.aar3617} {\bibfield  {journal} {\bibinfo  {journal}
  {Science}\ }\textbf {\bibinfo {volume} {360}},\ \bibinfo {pages} {1218}
  (\bibinfo {year} {2018})}\BibitemShut {NoStop}%
\bibitem [{\citenamefont {Zhou}\ \emph {et~al.}(2019)\citenamefont {Zhou},
  \citenamefont {Wang}, \citenamefont {Osterhoudt}, \citenamefont
  {Lampen-Kelley}, \citenamefont {Mandrus}, \citenamefont {He}, \citenamefont
  {Burch},\ and\ \citenamefont {Henriksen}}]{zhou2019possible}%
  \BibitemOpen
  \bibfield  {author} {\bibinfo {author} {\bibfnamefont {B.}~\bibnamefont
  {Zhou}}, \bibinfo {author} {\bibfnamefont {Y.}~\bibnamefont {Wang}}, \bibinfo
  {author} {\bibfnamefont {G.~B.}\ \bibnamefont {Osterhoudt}}, \bibinfo
  {author} {\bibfnamefont {P.}~\bibnamefont {Lampen-Kelley}}, \bibinfo {author}
  {\bibfnamefont {D.}~\bibnamefont {Mandrus}}, \bibinfo {author} {\bibfnamefont
  {R.}~\bibnamefont {He}}, \bibinfo {author} {\bibfnamefont {K.~S.}\
  \bibnamefont {Burch}}, \ and\ \bibinfo {author} {\bibfnamefont {E.~A.}\
  \bibnamefont {Henriksen}},\ }\href
  {https://www.sciencedirect.com/science/article/abs/pii/S0022369717315408}
  {\bibfield  {journal} {\bibinfo  {journal} {Journal of Physics and Chemistry
  of Solids}\ }\textbf {\bibinfo {volume} {128}},\ \bibinfo {pages} {291}
  (\bibinfo {year} {2019})}\BibitemShut {NoStop}%
\bibitem [{\citenamefont {Mashhadi}\ \emph {et~al.}(2019)\citenamefont
  {Mashhadi}, \citenamefont {Kim}, \citenamefont {Kim}, \citenamefont {Weber},
  \citenamefont {Taniguchi}, \citenamefont {Watanabe}, \citenamefont {Park},
  \citenamefont {Lotsch}, \citenamefont {Smet}, \citenamefont {Burghard} \emph
  {et~al.}}]{mashhadi2019spin}%
  \BibitemOpen
  \bibfield  {author} {\bibinfo {author} {\bibfnamefont {S.}~\bibnamefont
  {Mashhadi}}, \bibinfo {author} {\bibfnamefont {Y.}~\bibnamefont {Kim}},
  \bibinfo {author} {\bibfnamefont {J.}~\bibnamefont {Kim}}, \bibinfo {author}
  {\bibfnamefont {D.}~\bibnamefont {Weber}}, \bibinfo {author} {\bibfnamefont
  {T.}~\bibnamefont {Taniguchi}}, \bibinfo {author} {\bibfnamefont
  {K.}~\bibnamefont {Watanabe}}, \bibinfo {author} {\bibfnamefont
  {N.}~\bibnamefont {Park}}, \bibinfo {author} {\bibfnamefont {B.}~\bibnamefont
  {Lotsch}}, \bibinfo {author} {\bibfnamefont {J.~H.}\ \bibnamefont {Smet}},
  \bibinfo {author} {\bibfnamefont {M.}~\bibnamefont {Burghard}},  \emph
  {et~al.},\ }\href {https://pubs.acs.org/doi/abs/10.1021/acs.nanolett.9b01691}
  {\bibfield  {journal} {\bibinfo  {journal} {Nano Letters}\ }\textbf {\bibinfo
  {volume} {19}},\ \bibinfo {pages} {4659} (\bibinfo {year}
  {2019})}\BibitemShut {NoStop}%
\bibitem [{\citenamefont {Knolle}\ \emph {et~al.}(2014)\citenamefont {Knolle},
  \citenamefont {Kovrizhin}, \citenamefont {Chalker},\ and\ \citenamefont
  {Moessner}}]{KnolleMoessner2014}%
  \BibitemOpen
  \bibfield  {author} {\bibinfo {author} {\bibfnamefont {J.}~\bibnamefont
  {Knolle}}, \bibinfo {author} {\bibfnamefont {D.}~\bibnamefont {Kovrizhin}},
  \bibinfo {author} {\bibfnamefont {J.}~\bibnamefont {Chalker}}, \ and\
  \bibinfo {author} {\bibfnamefont {R.}~\bibnamefont {Moessner}},\ }\href
  {https://journals.aps.org/prl/abstract/10.1103/PhysRevLett.112.207203}
  {\bibfield  {journal} {\bibinfo  {journal} {Physical Review Letters}\
  }\textbf {\bibinfo {volume} {112}},\ \bibinfo {pages} {207203} (\bibinfo
  {year} {2014})}\BibitemShut {NoStop}%
\bibitem [{\citenamefont {Seifert}\ \emph {et~al.}(2018)\citenamefont
  {Seifert}, \citenamefont {Meng},\ and\ \citenamefont
  {Vojta}}]{SeifertVojta2018}%
  \BibitemOpen
  \bibfield  {author} {\bibinfo {author} {\bibfnamefont {U.~F.}\ \bibnamefont
  {Seifert}}, \bibinfo {author} {\bibfnamefont {T.}~\bibnamefont {Meng}}, \
  and\ \bibinfo {author} {\bibfnamefont {M.}~\bibnamefont {Vojta}},\ }\href
  {https://journals.aps.org/prb/abstract/10.1103/PhysRevB.97.085118} {\bibfield
   {journal} {\bibinfo  {journal} {Physical Review B}\ }\textbf {\bibinfo
  {volume} {97}},\ \bibinfo {pages} {085118} (\bibinfo {year}
  {2018})}\BibitemShut {NoStop}%
\bibitem [{\citenamefont {Choi}\ \emph {et~al.}(2018)\citenamefont {Choi},
  \citenamefont {Klein}, \citenamefont {Rosch},\ and\ \citenamefont
  {Kim}}]{ChoiKim2018}%
  \BibitemOpen
  \bibfield  {author} {\bibinfo {author} {\bibfnamefont {W.}~\bibnamefont
  {Choi}}, \bibinfo {author} {\bibfnamefont {P.~W.}\ \bibnamefont {Klein}},
  \bibinfo {author} {\bibfnamefont {A.}~\bibnamefont {Rosch}}, \ and\ \bibinfo
  {author} {\bibfnamefont {Y.~B.}\ \bibnamefont {Kim}},\ }\href
  {https://journals.aps.org/prb/abstract/10.1103/PhysRevB.98.155123} {\bibfield
   {journal} {\bibinfo  {journal} {Physical Review B}\ }\textbf {\bibinfo
  {volume} {98}},\ \bibinfo {pages} {155123} (\bibinfo {year}
  {2018})}\BibitemShut {NoStop}%
\bibitem [{\citenamefont {Winter}\ \emph {et~al.}(2016)\citenamefont {Winter},
  \citenamefont {Li}, \citenamefont {Jeschke},\ and\ \citenamefont
  {Valent{\'\i}}}]{winter2016challenges}%
  \BibitemOpen
  \bibfield  {author} {\bibinfo {author} {\bibfnamefont {S.~M.}\ \bibnamefont
  {Winter}}, \bibinfo {author} {\bibfnamefont {Y.}~\bibnamefont {Li}}, \bibinfo
  {author} {\bibfnamefont {H.~O.}\ \bibnamefont {Jeschke}}, \ and\ \bibinfo
  {author} {\bibfnamefont {R.}~\bibnamefont {Valent{\'\i}}},\ }\href
  {https://journals.aps.org/prb/abstract/10.1103/PhysRevB.93.214431} {\bibfield
   {journal} {\bibinfo  {journal} {Physical Review B}\ }\textbf {\bibinfo
  {volume} {93}},\ \bibinfo {pages} {214431} (\bibinfo {year}
  {2016})}\BibitemShut {NoStop}%
\bibitem [{\citenamefont {Baskaran}\ \emph {et~al.}(2007)\citenamefont
  {Baskaran}, \citenamefont {Mandal},\ and\ \citenamefont
  {Shankar}}]{Baskaran2007}%
  \BibitemOpen
  \bibfield  {author} {\bibinfo {author} {\bibfnamefont {G.}~\bibnamefont
  {Baskaran}}, \bibinfo {author} {\bibfnamefont {S.}~\bibnamefont {Mandal}}, \
  and\ \bibinfo {author} {\bibfnamefont {R.}~\bibnamefont {Shankar}},\ }\href
  {\doibase 10.1103/PhysRevLett.98.247201} {\bibfield  {journal} {\bibinfo
  {journal} {Phys. Rev. Lett.}\ }\textbf {\bibinfo {volume} {98}},\ \bibinfo
  {pages} {247201} (\bibinfo {year} {2007})}\BibitemShut {NoStop}%
\bibitem [{\citenamefont {Fern\'andez-Rossier}(2009)}]{rossier2009theo}%
  \BibitemOpen
  \bibfield  {author} {\bibinfo {author} {\bibfnamefont {J.}~\bibnamefont
  {Fern\'andez-Rossier}},\ }\href {\doibase 10.1103/PhysRevLett.102.256802}
  {\bibfield  {journal} {\bibinfo  {journal} {Phys. Rev. Lett.}\ }\textbf
  {\bibinfo {volume} {102}},\ \bibinfo {pages} {256802} (\bibinfo {year}
  {2009})}\BibitemShut {NoStop}%
\bibitem [{\citenamefont {Fransson}\ \emph {et~al.}(2010)\citenamefont
  {Fransson}, \citenamefont {Eriksson},\ and\ \citenamefont
  {Balatsky}}]{Balatsky2010theo}%
  \BibitemOpen
  \bibfield  {author} {\bibinfo {author} {\bibfnamefont {J.}~\bibnamefont
  {Fransson}}, \bibinfo {author} {\bibfnamefont {O.}~\bibnamefont {Eriksson}},
  \ and\ \bibinfo {author} {\bibfnamefont {A.~V.}\ \bibnamefont {Balatsky}},\
  }\href {\doibase 10.1103/PhysRevB.81.115454} {\bibfield  {journal} {\bibinfo
  {journal} {Phys. Rev. B}\ }\textbf {\bibinfo {volume} {81}},\ \bibinfo
  {pages} {115454} (\bibinfo {year} {2010})}\BibitemShut {NoStop}%
\bibitem [{\citenamefont {Rantner}\ and\ \citenamefont
  {Wen}(2001)}]{RantnerWen2001}%
  \BibitemOpen
  \bibfield  {author} {\bibinfo {author} {\bibfnamefont {W.}~\bibnamefont
  {Rantner}}\ and\ \bibinfo {author} {\bibfnamefont {X.-G.}\ \bibnamefont
  {Wen}},\ }\href {\doibase 10.1103/PhysRevLett.86.3871} {\bibfield  {journal}
  {\bibinfo  {journal} {Phys. Rev. Lett.}\ }\textbf {\bibinfo {volume} {86}},\
  \bibinfo {pages} {3871} (\bibinfo {year} {2001})}\BibitemShut {NoStop}%
\bibitem [{Sup()}]{SuppMat}%
  \BibitemOpen
  \href@noop {} {}\bibinfo {note} {See supplementary materials to this
  publication.}\BibitemShut {Stop}%
\bibitem [{\citenamefont {Wang}\ \emph {et~al.}(2017)\citenamefont {Wang},
  \citenamefont {Reschke}, \citenamefont {H{\"u}vonen}, \citenamefont {Do},
  \citenamefont {Choi}, \citenamefont {Gensch}, \citenamefont {Nagel},
  \citenamefont {R{\~o}{\~o}m},\ and\ \citenamefont
  {Loidl}}]{wang2017magnetic}%
  \BibitemOpen
  \bibfield  {author} {\bibinfo {author} {\bibfnamefont {Z.}~\bibnamefont
  {Wang}}, \bibinfo {author} {\bibfnamefont {S.}~\bibnamefont {Reschke}},
  \bibinfo {author} {\bibfnamefont {D.}~\bibnamefont {H{\"u}vonen}}, \bibinfo
  {author} {\bibfnamefont {S.-H.}\ \bibnamefont {Do}}, \bibinfo {author}
  {\bibfnamefont {K.-Y.}\ \bibnamefont {Choi}}, \bibinfo {author}
  {\bibfnamefont {M.}~\bibnamefont {Gensch}}, \bibinfo {author} {\bibfnamefont
  {U.}~\bibnamefont {Nagel}}, \bibinfo {author} {\bibfnamefont
  {T.}~\bibnamefont {R{\~o}{\~o}m}}, \ and\ \bibinfo {author} {\bibfnamefont
  {A.}~\bibnamefont {Loidl}},\ }\href
  {https://journals.aps.org/prl/abstract/10.1103/PhysRevLett.119.227202}
  {\bibfield  {journal} {\bibinfo  {journal} {Physical review letters}\
  }\textbf {\bibinfo {volume} {119}},\ \bibinfo {pages} {227202} (\bibinfo
  {year} {2017})}\BibitemShut {NoStop}%
\bibitem [{\citenamefont {Little}\ \emph {et~al.}(2017)\citenamefont {Little},
  \citenamefont {Wu}, \citenamefont {Lampen-Kelley}, \citenamefont {Banerjee},
  \citenamefont {Patankar}, \citenamefont {Rees}, \citenamefont {Bridges},
  \citenamefont {Yan}, \citenamefont {Mandrus}, \citenamefont {Nagler} \emph
  {et~al.}}]{little2017antiferromagnetic}%
  \BibitemOpen
  \bibfield  {author} {\bibinfo {author} {\bibfnamefont {A.}~\bibnamefont
  {Little}}, \bibinfo {author} {\bibfnamefont {L.}~\bibnamefont {Wu}}, \bibinfo
  {author} {\bibfnamefont {P.}~\bibnamefont {Lampen-Kelley}}, \bibinfo {author}
  {\bibfnamefont {A.}~\bibnamefont {Banerjee}}, \bibinfo {author}
  {\bibfnamefont {S.}~\bibnamefont {Patankar}}, \bibinfo {author}
  {\bibfnamefont {D.}~\bibnamefont {Rees}}, \bibinfo {author} {\bibfnamefont
  {C.}~\bibnamefont {Bridges}}, \bibinfo {author} {\bibfnamefont {J.-Q.}\
  \bibnamefont {Yan}}, \bibinfo {author} {\bibfnamefont {D.}~\bibnamefont
  {Mandrus}}, \bibinfo {author} {\bibfnamefont {S.}~\bibnamefont {Nagler}},
  \emph {et~al.},\ }\href
  {https://journals.aps.org/prl/abstract/10.1103/PhysRevLett.119.227201}
  {\bibfield  {journal} {\bibinfo  {journal} {Physical review letters}\
  }\textbf {\bibinfo {volume} {119}},\ \bibinfo {pages} {227201} (\bibinfo
  {year} {2017})}\BibitemShut {NoStop}%
\bibitem [{\citenamefont {Murphy}\ \emph {et~al.}(1995)\citenamefont {Murphy},
  \citenamefont {Eisenstein}, \citenamefont {Pfeiffer},\ and\ \citenamefont
  {West}}]{MurphyWest1995}%
  \BibitemOpen
  \bibfield  {author} {\bibinfo {author} {\bibfnamefont {S.~Q.}\ \bibnamefont
  {Murphy}}, \bibinfo {author} {\bibfnamefont {J.~P.}\ \bibnamefont
  {Eisenstein}}, \bibinfo {author} {\bibfnamefont {L.~N.}\ \bibnamefont
  {Pfeiffer}}, \ and\ \bibinfo {author} {\bibfnamefont {K.~W.}\ \bibnamefont
  {West}},\ }\href {\doibase 10.1103/PhysRevB.52.14825} {\bibfield  {journal}
  {\bibinfo  {journal} {Phys. Rev. B}\ }\textbf {\bibinfo {volume} {52}},\
  \bibinfo {pages} {14825} (\bibinfo {year} {1995})}\BibitemShut {NoStop}%
\bibitem [{\citenamefont {Biswas}\ \emph {et~al.}(2019)\citenamefont {Biswas},
  \citenamefont {Li}, \citenamefont {Winter}, \citenamefont {Knolle},\ and\
  \citenamefont {Valent\'{\i}}}]{BiswasValenti2018}%
  \BibitemOpen
  \bibfield  {author} {\bibinfo {author} {\bibfnamefont {S.}~\bibnamefont
  {Biswas}}, \bibinfo {author} {\bibfnamefont {Y.}~\bibnamefont {Li}}, \bibinfo
  {author} {\bibfnamefont {S.~M.}\ \bibnamefont {Winter}}, \bibinfo {author}
  {\bibfnamefont {J.}~\bibnamefont {Knolle}}, \ and\ \bibinfo {author}
  {\bibfnamefont {R.}~\bibnamefont {Valent\'{\i}}},\ }\href {\doibase
  10.1103/PhysRevLett.123.237201} {\bibfield  {journal} {\bibinfo  {journal}
  {Phys. Rev. Lett.}\ }\textbf {\bibinfo {volume} {123}},\ \bibinfo {pages}
  {237201} (\bibinfo {year} {2019})}\BibitemShut {NoStop}%
\bibitem [{\citenamefont {Chen}\ and\ \citenamefont
  {Lado}(2020)}]{chen2020impurity}%
  \BibitemOpen
  \bibfield  {author} {\bibinfo {author} {\bibfnamefont {G.}~\bibnamefont
  {Chen}}\ and\ \bibinfo {author} {\bibfnamefont {J.}~\bibnamefont {Lado}},\
  }\href {https://arxiv.org/abs/2005.06896} {\bibfield  {journal} {\bibinfo
  {journal} {arXiv preprint arXiv:2005.06896}\ } (\bibinfo {year}
  {2020})}\BibitemShut {NoStop}%
\bibitem [{\citenamefont {Feldmeier}\ \emph {et~al.}(2020)\citenamefont
  {Feldmeier}, \citenamefont {Natori}, \citenamefont {Knap},\ and\
  \citenamefont {Knolle}}]{feldmeier2020local}%
  \BibitemOpen
  \bibfield  {author} {\bibinfo {author} {\bibfnamefont {J.}~\bibnamefont
  {Feldmeier}}, \bibinfo {author} {\bibfnamefont {W.}~\bibnamefont {Natori}},
  \bibinfo {author} {\bibfnamefont {M.}~\bibnamefont {Knap}}, \ and\ \bibinfo
  {author} {\bibfnamefont {J.}~\bibnamefont {Knolle}},\ }\href
  {https://arxiv.org/abs/2007.07912} {\bibfield  {journal} {\bibinfo  {journal}
  {arXiv preprint arXiv:2007.07912}\ } (\bibinfo {year} {2020})}\BibitemShut
  {NoStop}%
\bibitem [{\citenamefont {Law}\ and\ \citenamefont {Lee}(2017)}]{law20171t}%
  \BibitemOpen
  \bibfield  {author} {\bibinfo {author} {\bibfnamefont {K.~T.}\ \bibnamefont
  {Law}}\ and\ \bibinfo {author} {\bibfnamefont {P.~A.}\ \bibnamefont {Lee}},\
  }\href {https://www.pnas.org/content/114/27/6996} {\bibfield  {journal}
  {\bibinfo  {journal} {Proceedings of the National Academy of Sciences}\
  }\textbf {\bibinfo {volume} {114}},\ \bibinfo {pages} {6996} (\bibinfo {year}
  {2017})}\BibitemShut {NoStop}%
\bibitem [{\citenamefont {Kratochvilova}\ \emph {et~al.}(2017)\citenamefont
  {Kratochvilova}, \citenamefont {Hillier}, \citenamefont {Wildes},
  \citenamefont {Wang}, \citenamefont {Cheong},\ and\ \citenamefont
  {Park}}]{KratochvilovaPark2017}%
  \BibitemOpen
  \bibfield  {author} {\bibinfo {author} {\bibfnamefont {M.}~\bibnamefont
  {Kratochvilova}}, \bibinfo {author} {\bibfnamefont {A.~D.}\ \bibnamefont
  {Hillier}}, \bibinfo {author} {\bibfnamefont {A.~R.}\ \bibnamefont {Wildes}},
  \bibinfo {author} {\bibfnamefont {L.}~\bibnamefont {Wang}}, \bibinfo {author}
  {\bibfnamefont {S.-W.}\ \bibnamefont {Cheong}}, \ and\ \bibinfo {author}
  {\bibfnamefont {J.-G.}\ \bibnamefont {Park}},\ }\href
  {https://www.nature.com/articles/s41535-017-0048-1} {\bibfield  {journal}
  {\bibinfo  {journal} {npj Quantum Materials}\ }\textbf {\bibinfo {volume}
  {2}},\ \bibinfo {pages} {1} (\bibinfo {year} {2017})}\BibitemShut {NoStop}%
\bibitem [{\citenamefont {Nakata}\ \emph {et~al.}(2018)\citenamefont {Nakata},
  \citenamefont {Yoshizawa}, \citenamefont {Sugawara}, \citenamefont {Umemoto},
  \citenamefont {Takahashi},\ and\ \citenamefont {Sato}}]{nakata2018selective}%
  \BibitemOpen
  \bibfield  {author} {\bibinfo {author} {\bibfnamefont {Y.}~\bibnamefont
  {Nakata}}, \bibinfo {author} {\bibfnamefont {T.}~\bibnamefont {Yoshizawa}},
  \bibinfo {author} {\bibfnamefont {K.}~\bibnamefont {Sugawara}}, \bibinfo
  {author} {\bibfnamefont {Y.}~\bibnamefont {Umemoto}}, \bibinfo {author}
  {\bibfnamefont {T.}~\bibnamefont {Takahashi}}, \ and\ \bibinfo {author}
  {\bibfnamefont {T.}~\bibnamefont {Sato}},\ }\href
  {https://pubs.acs.org/doi/abs/10.1021/acsanm.8b00184} {\bibfield  {journal}
  {\bibinfo  {journal} {ACS Applied Nano Materials}\ }\textbf {\bibinfo
  {volume} {1}},\ \bibinfo {pages} {1456} (\bibinfo {year} {2018})}\BibitemShut
  {NoStop}%
\bibitem [{\citenamefont {Lin}\ \emph {et~al.}(2018)\citenamefont {Lin},
  \citenamefont {Huang}, \citenamefont {Zhao}, \citenamefont {Lian},
  \citenamefont {Duan}, \citenamefont {Chen},\ and\ \citenamefont
  {Ji}}]{lin2018growth}%
  \BibitemOpen
  \bibfield  {author} {\bibinfo {author} {\bibfnamefont {H.}~\bibnamefont
  {Lin}}, \bibinfo {author} {\bibfnamefont {W.}~\bibnamefont {Huang}}, \bibinfo
  {author} {\bibfnamefont {K.}~\bibnamefont {Zhao}}, \bibinfo {author}
  {\bibfnamefont {C.}~\bibnamefont {Lian}}, \bibinfo {author} {\bibfnamefont
  {W.}~\bibnamefont {Duan}}, \bibinfo {author} {\bibfnamefont {X.}~\bibnamefont
  {Chen}}, \ and\ \bibinfo {author} {\bibfnamefont {S.-H.}\ \bibnamefont
  {Ji}},\ }\href {https://link.springer.com/article/10.1007/s12274-018-2054-4}
  {\bibfield  {journal} {\bibinfo  {journal} {Nano Research}\ }\textbf
  {\bibinfo {volume} {11}},\ \bibinfo {pages} {4722} (\bibinfo {year}
  {2018})}\BibitemShut {NoStop}%
\bibitem [{\citenamefont {Lin}\ \emph {et~al.}(2020)\citenamefont {Lin},
  \citenamefont {Huang}, \citenamefont {Zhao}, \citenamefont {Qiao},
  \citenamefont {Liu}, \citenamefont {Wu}, \citenamefont {Chen},\ and\
  \citenamefont {Ji}}]{lin2020scanning}%
  \BibitemOpen
  \bibfield  {author} {\bibinfo {author} {\bibfnamefont {H.}~\bibnamefont
  {Lin}}, \bibinfo {author} {\bibfnamefont {W.}~\bibnamefont {Huang}}, \bibinfo
  {author} {\bibfnamefont {K.}~\bibnamefont {Zhao}}, \bibinfo {author}
  {\bibfnamefont {S.}~\bibnamefont {Qiao}}, \bibinfo {author} {\bibfnamefont
  {Z.}~\bibnamefont {Liu}}, \bibinfo {author} {\bibfnamefont {J.}~\bibnamefont
  {Wu}}, \bibinfo {author} {\bibfnamefont {X.}~\bibnamefont {Chen}}, \ and\
  \bibinfo {author} {\bibfnamefont {S.-H.}\ \bibnamefont {Ji}},\ }\href
  {https://link.springer.com/article/10.1007/s12274-019-2584-4} {\bibfield
  {journal} {\bibinfo  {journal} {Nano Research}\ }\textbf {\bibinfo {volume}
  {13}},\ \bibinfo {pages} {133} (\bibinfo {year} {2020})}\BibitemShut
  {NoStop}%
\bibitem [{\citenamefont {Chen}\ \emph {et~al.}(2020)\citenamefont {Chen},
  \citenamefont {Ruan}, \citenamefont {Wu}, \citenamefont {Tang}, \citenamefont
  {Ryu}, \citenamefont {Tsai}, \citenamefont {Lee}, \citenamefont {Kahn},
  \citenamefont {Liou}, \citenamefont {Jia}, \citenamefont {Albertini},
  \citenamefont {Xiong}, \citenamefont {Jia}, \citenamefont {Liu},
  \citenamefont {Sobota}, \citenamefont {Liu}, \citenamefont {Moore},
  \citenamefont {Shen}, \citenamefont {Louie}, \citenamefont {Mo},\ and\
  \citenamefont {Crommie}}]{chen2020mott}%
  \BibitemOpen
  \bibfield  {author} {\bibinfo {author} {\bibfnamefont {Y.}~\bibnamefont
  {Chen}}, \bibinfo {author} {\bibfnamefont {W.}~\bibnamefont {Ruan}}, \bibinfo
  {author} {\bibfnamefont {M.}~\bibnamefont {Wu}}, \bibinfo {author}
  {\bibfnamefont {S.}~\bibnamefont {Tang}}, \bibinfo {author} {\bibfnamefont
  {H.}~\bibnamefont {Ryu}}, \bibinfo {author} {\bibfnamefont {H.-Z.}\
  \bibnamefont {Tsai}}, \bibinfo {author} {\bibfnamefont {R.}~\bibnamefont
  {Lee}}, \bibinfo {author} {\bibfnamefont {S.}~\bibnamefont {Kahn}}, \bibinfo
  {author} {\bibfnamefont {F.}~\bibnamefont {Liou}}, \bibinfo {author}
  {\bibfnamefont {C.}~\bibnamefont {Jia}}, \bibinfo {author} {\bibfnamefont
  {O.~R.}\ \bibnamefont {Albertini}}, \bibinfo {author} {\bibfnamefont
  {H.}~\bibnamefont {Xiong}}, \bibinfo {author} {\bibfnamefont
  {T.}~\bibnamefont {Jia}}, \bibinfo {author} {\bibfnamefont {Z.}~\bibnamefont
  {Liu}}, \bibinfo {author} {\bibfnamefont {J.~A.}\ \bibnamefont {Sobota}},
  \bibinfo {author} {\bibfnamefont {A.~Y.}\ \bibnamefont {Liu}}, \bibinfo
  {author} {\bibfnamefont {J.~E.}\ \bibnamefont {Moore}}, \bibinfo {author}
  {\bibfnamefont {Z.-X.}\ \bibnamefont {Shen}}, \bibinfo {author}
  {\bibfnamefont {S.~G.}\ \bibnamefont {Louie}}, \bibinfo {author}
  {\bibfnamefont {S.-K.}\ \bibnamefont {Mo}}, \ and\ \bibinfo {author}
  {\bibfnamefont {M.~F.}\ \bibnamefont {Crommie}},\ }\href {\doibase
  10.1038/s41567-019-0744-9} {\bibfield  {journal} {\bibinfo  {journal} {Nature
  Physics}\ }\textbf {\bibinfo {volume} {16}},\ \bibinfo {pages} {218}
  (\bibinfo {year} {2020})}\BibitemShut {NoStop}%
\bibitem [{\citenamefont {Boese}\ \emph {et~al.}(2001)\citenamefont {Boese},
  \citenamefont {Governale}, \citenamefont {Rosch},\ and\ \citenamefont
  {Z\"ulicke}}]{BoeseZuelicke2001}%
  \BibitemOpen
  \bibfield  {author} {\bibinfo {author} {\bibfnamefont {D.}~\bibnamefont
  {Boese}}, \bibinfo {author} {\bibfnamefont {M.}~\bibnamefont {Governale}},
  \bibinfo {author} {\bibfnamefont {A.}~\bibnamefont {Rosch}}, \ and\ \bibinfo
  {author} {\bibfnamefont {U.}~\bibnamefont {Z\"ulicke}},\ }\href {\doibase
  10.1103/PhysRevB.64.085315} {\bibfield  {journal} {\bibinfo  {journal} {Phys.
  Rev. B}\ }\textbf {\bibinfo {volume} {64}},\ \bibinfo {pages} {085315}
  (\bibinfo {year} {2001})}\BibitemShut {NoStop}%
\bibitem [{\citenamefont {Rammer}\ and\ \citenamefont
  {Smith}(1986)}]{RammerSmith}%
  \BibitemOpen
  \bibfield  {author} {\bibinfo {author} {\bibfnamefont {J.}~\bibnamefont
  {Rammer}}\ and\ \bibinfo {author} {\bibfnamefont {H.}~\bibnamefont {Smith}},\
  }\href {\doibase 10.1103/RevModPhys.58.323} {\bibfield  {journal} {\bibinfo
  {journal} {Rev. Mod. Phys.}\ }\textbf {\bibinfo {volume} {58}},\ \bibinfo
  {pages} {323} (\bibinfo {year} {1986})}\BibitemShut {NoStop}%
\bibitem [{\citenamefont {Horiguchi}(1972)}]{Horiguchi1972}%
  \BibitemOpen
  \bibfield  {author} {\bibinfo {author} {\bibfnamefont {T.}~\bibnamefont
  {Horiguchi}},\ }\href {https://aip.scitation.org/doi/abs/10.1063/1.1666155}
  {\bibfield  {journal} {\bibinfo  {journal} {Journal of Mathematical Physics}\
  }\textbf {\bibinfo {volume} {13}},\ \bibinfo {pages} {1411} (\bibinfo {year}
  {1972})}\BibitemShut {NoStop}%
\bibitem [{\citenamefont {Auerbach}(2012)}]{AuerbachBook}%
  \BibitemOpen
  \bibfield  {author} {\bibinfo {author} {\bibfnamefont {A.}~\bibnamefont
  {Auerbach}},\ }\href {https://books.google.de/books?id=d-sHCAAAQBAJ} {\emph
  {\bibinfo {title} {Interacting Electrons and Quantum Magnetism}}},\ Graduate
  Texts in Contemporary Physics\ (\bibinfo  {publisher} {Springer New York},\
  \bibinfo {year} {2012})\BibitemShut {NoStop}%
\end{thebibliography}%

\newpage
\clearpage

\setcounter{equation}{0}
\setcounter{figure}{0}
\setcounter{section}{0}
\setcounter{table}{0}
\setcounter{page}{1}
\makeatletter
\renewcommand{\theequation}{S\arabic{equation}}
\renewcommand{\thesection}{S\arabic{section}}
\renewcommand{\thefigure}{S\arabic{figure}}

\begin{widetext}
\begin{center}
Supplementary materials on \\
\textbf{"Tunneling spectroscopy of quantum spin liquids"}\\
Elio J. K\"onig$^{1}$, Mallika T. Randeria$^{2}$, Berthold J\"ack$^{3}$\\ 
$^{1}$\textit{Department of Physics and Astronomy, Center for Materials Theory, Rutgers University, Piscataway, NJ 08854}
$^{2}$\textit{Department of Physics, Massachusetts Institute of Technology, Cambridge, Massachusetts 02139, USA}\\
$^3$\textit{Princeton University, Joseph Henry Laboratory at the Department of Physics, Princeton, NJ 08544, USA}

\end{center}

\end{widetext}

These supplementary materials contain a derivation of the tunneling current, Sec.~\ref{tuncurr}, a summary of the Kitaev spin liquid, Sec.~\ref{sec:KitaevModel} and a derivation of the tunneling response for a monolayer 2D N\'eel antiferromagnet, Sec.~\ref{sec:AFM}.

\section{Derivation of tunneling current}
\label{tuncurr}

In this section, we present formal details on the derivation of the tunneling current for AC linear response and DC non-linear response in the case of a point contact or a planar tunnel junction.

In this appendix, we set $e^2 = \hbar = 1$ and we use the notation $eV = \mu_2 - \mu_1 \equiv \mu_{21}$.

\subsection{Evaluation of response functions}
We use the following expression for the DC nonlinear  current accross the junction
\begin{subequations}
\begin{equation}
I = \langle I_{\rm tot}(t) \rangle = - i \int dt'\Theta(t-t') \sum_{\v x, \v x'}\langle [ \hat I_{\v x}(t), \hat T_{\v x'}(t') ] \rangle,
\end{equation}
and for the AC conductance (in linear response)
\begin{equation}
G(\Omega) = \int dt'\Theta(t-t') \frac{e^{i \Omega(t- t')}}{\Omega} \sum_{\v x, \v x'}\langle \left [\hat I_{\v x}(t),\hat I_{\v x'}(t') \right ] \rangle_{eV = 0}.
\end{equation}
\label{eq:BasicResponseFunctions}
\end{subequations}

Here, we introduced the local  current operator
\begin{subequations}
\begin{align}
\hat I_{\v x}(t) &=  \left(t_{\v x} \delta_{\sigma \sigma'} + J_{\v x} \vec \sigma_{\sigma, \sigma'} \cdot \hat{\vec S}({\bf x}) \right)\notag \\
&\left [i c^\dagger_{{\bf x} 1 \sigma} c_{{\bf x} 2 \sigma'}e^{i \mu_{21}t} -ic^\dagger_{{\bf x} 2 \sigma} c_{{\bf x} 1 \sigma'} e^{-i \mu_{21}t}  \right ]\label{eq:I}
\end{align}
as well as the local hopping term
\begin{align}
\hat T_{\v x}(t) &=  \left(t_{\v x} \delta_{\sigma \sigma'} + J_{\v x} \vec \sigma_{\sigma, \sigma'} \cdot \hat{\vec S}({\bf x}) \right)\notag \\
&\left [ c^\dagger_{{\bf x} 1 \sigma} c_{{\bf x} 2 \sigma'}e^{i \mu_{21}t} +c^\dagger_{{\bf x} 2 \sigma} c_{{\bf x} 1 \sigma'}e^{-i \mu_{21}t}  \right ].
\end{align}
\end{subequations}

The time evolution in Eq.~\eqref{eq:BasicResponseFunctions} is with respect to $H_{\rm leads} + H_{\rm QSL}$ of Eq.~\eqref{eq:H0} of the main text.
We use 
\begin{widetext}
\begin{subequations}
\begin{align}
\langle [\hat I_{\v x}(t),\hat T_{\v x'}(t') ]\rangle &= 2i[t_{\v x} t_{\v x'} + iJ_{\v x} J_{\v x'} C^>(\v x, \v x'; t, t')][i\Pi_{1^\dagger 2}^>(\v x, \v x'; t, t')e^{i \mu_{21}(t-t')} - 1 \leftrightarrow  2] - >\, \rightarrow \, < \notag \\
&\doteq -2 t_{\v x} t_{\v x'}[\Pi_{1^\dagger 2}^+(\v x, \v x'; t, t')e^{i \mu_{21}(t-t')} - 1 \leftrightarrow  2] \notag \\
&-i J_{\v x} J_{\v x'} \left \lbrace C^K(\v x, \v x'; t, t')][\Pi^+_{1^\dagger 2}(\v x, \v x'; t, t')e^{i \mu_{21}(t-t')} - 1 \leftrightarrow  2] \right. \notag \\
&\left .+ C^+(\v x, \v x'; t, t')][\Pi_{1^\dagger 2}^K(\v x, \v x'; t,t')e^{i \mu_{21}(t-t')} - 1 \leftrightarrow  2]\right \rbrace,\\
\langle [\hat I_{\v x}(t),\hat I_{\v x'}(t') ]\rangle_{eV = 0} &= 2[t_{\v x} t_{\v x'} + iJ_{\v x} J_{\v x'} C^>(\v x, \v x'; t, t')][i\Pi_{1^\dagger 2}^>(\v x, \v x'; t, t') + 1 \leftrightarrow  2] - >\, \rightarrow \, < \notag \\
&\doteq 2i t_{\v x} t_{\v x'}[\Pi_{1^\dagger 2}^+(\v x, \v x'; t, t') + 1 \leftrightarrow  2] \notag \\
&- J_{\v x} J_{\v x'} \left \lbrace C^K(\v x, \v x'; t, t')][\Pi^+_{1^\dagger 2}(\v x, \v x'; t, t') + 1 \leftrightarrow  2] + C^+(\v x, \v x'; t, t')][\Pi_{1^\dagger 2}^K(\v x, \v x'; t,t') + 1 \leftrightarrow  2]\right \rbrace.
\end{align}
\end{subequations}
\end{widetext}
At the `$\doteq$' sign, we used that the commutators are multiplied by $\Theta(t - t')$. We further employed the standard notation for greater and lesser~\cite{RammerSmith} correlators: 
\begin{subequations}
\begin{align}
\Pi^<_{1^\dagger 2}(\v x, \v x'; t,t') &= - i \langle [c^\dagger_{\v x' 2} c_{\v x' 1}](t') [c^\dagger_{\v x 1} c_{\v x 2}](t) \rangle ,\\
\Pi^>_{1^\dagger 2}(\v x, \v x'; t,t') &= - i \langle [c^\dagger_{\v x 1} c_{\v x 2}](t) [c^\dagger_{\v x' 2} c_{\v x' 1}](t') \rangle,\\
C^<(\v x, \v x'; t,t') &= - i \left \langle \vec S(\v x',t') \cdot \vec S(\v x,t) \right \rangle,\\
C^>(\v x, \v x'; t,t')&=-i\left \langle \vec S(\v x,t) \cdot \vec S(\v x',t') \right \rangle .
\end{align}
\end{subequations}
Clearly, $C^<(\v x, \v x'; t,t')= C^>(\v x', \v x; t',t)$ and $\Pi^>_{2^\dagger 1}(\v x, \v x'; t, t') = 
\Pi^<_{1^\dagger 2} (\v x', \v x; t', t)$. We have suppressed the spin index of electronic operators (the present correlators are equal spin correlators). We furthermore used the following relationship to retarded (indicated by a $+$) and Keldysh Green's functions 
\begin{subequations}
\begin{align}
\Pi^+_{1^\dagger 2}(\v x, \v x'; t,t') &=  \Theta (t - t')\notag \\
&\times [\Pi^>_{1^\dagger 2}(\v x, \v x'; t,t')- \Pi^<_{1^\dagger 2}(\v x, \v x'; t,t')], \label{eq:PRet}\\
\Pi^K_{1^\dagger 2}(\v x, \v x'; t,t') &= \Pi^>_{1^\dagger 2}(\v x, \v x'; t,t')+ \Pi^<_{1^\dagger 2}(\v x, \v x'; t,t'),
\end{align}
\end{subequations}
and analogously for $C^{<,>,K,+}(\v x, \v x'; t,t')$.
It is convenient to Fourier transform all correlators in time/frequency space, such that
\begin{subequations}
\begin{align}
I &= i \sum_{\v x, \v x'} \Big \lbrace 2 t_{\v x}t_{\v x'}  \left [\Pi^+_{1^\dagger 2}(\v x, \v x'; \mu_{21}) - 1 \leftrightarrow  2 \right ]\notag\\
&+i \int \frac{d\omega}{2\pi} {J_{\v x} J_{\v x'}} \big [ C^K(\v x, \v x';\mu_{21} - \omega)\Pi^+_{1^\dagger 2}(\v x, \v x'; \omega) \notag\\
&+ C^+(\v x, \v x'; \mu_{21}- \omega)\Pi_{1^\dagger 2}^K(\v x, \v x'; \omega) - 1 \leftrightarrow  2 \big ] \Big \rbrace,\\
G(\Omega) &= - \sum_{\v x, \v x'} \Big \lbrace \frac{2 t_{\v x} t_{\v x'}}{\Omega i} \left [\Pi^+_{1^\dagger 2}(\v x, \v x'; \Omega) + 1 \leftrightarrow  2 \right ]\notag \\
&+ \int \frac{d\omega}{2\pi} \frac{J_{\v x} J_{\v x'}}{\Omega} \big [ C^K(\v x, \v x'; \Omega - \omega)\Pi^+_{1^\dagger 2}(\v x, \v x'; \omega) \notag\\
&+ C^+(\v x, \v x'; \Omega - \omega)\Pi_{1^\dagger 2}^K(\v x, \v x'; \omega) + 1 \leftrightarrow  2 \big ] \Big \rbrace.
\end{align}
\end{subequations}

\subsection{Equilibrium response}

We assume thermodynamic equilibrium of the spin system, which implies by the fluctuation dissipation theorem that 
\begin{equation}
C^K (\v x, \v x'; \omega) = 2i \coth\left (\frac{\omega}{2T} \right) \Im [C^+(\v x, \v x'; \omega)].
\end{equation}
An analogous expression holds for $\Pi^{K,+}$.
We can use the identity
\begin{align}
\frac{f(\epsilon')[1- f(\epsilon)] + \epsilon \leftrightarrow \epsilon'}{f(\epsilon') - f(\epsilon)} = \coth \left (\frac{\epsilon - \epsilon'}{2T} \right )
\end{align}
to reexpress
\begin{subequations}
\begin{align}
\Pi_{1^\dagger 2}^+(\v x, \v x'; \omega) &= \int (d\omega') \frac{1}{\omega - \omega' + i \eta} \label{eq:PiAs} \notag \\
& \int (dP) (dP') e^{i (\v p - \v p')(\v x - \v x')} \notag \\\
&(2\pi) \delta(\omega' - (\epsilon - \epsilon'))  A_2(\v p,\epsilon) A_1(\v p',\epsilon')\notag \\
& [f(\epsilon')-f(\epsilon)], \\
\Pi_{1^\dagger 2}^K(\v x, \v x'; \omega) &= 2i  \coth \left(\frac{\omega}{2T} \right) \Im [\Pi_{1^\dagger 2}^+(\v x, \v x'; \omega)],
\end{align}
\end{subequations}
where the limit $\eta \rightarrow 0$ is implied.
Here, $(dP) = d^2 p d \epsilon/(2\pi)^3$, $f(\epsilon) = [e^{\epsilon/T}+1]^{-1}$ is the Fermi-Dirac distribution and the spectral weight of conduction fermions in lead 1 (2) is denoted $A_{1 (2)}(\v p, \epsilon)$. For simplicity we use the same quadratic dispersion and scattering rate for both leads
\begin{subequations}
\begin{align}
A_2(\epsilon) = A_1(\epsilon) &=  \frac{2\Gamma}{(\epsilon - \frac{p^2}{2m} + \mu)^2 + \Gamma^2}.
\end{align}
\end{subequations}
where $\Gamma$, is the phenomenological decay rate. (In Eq.~\eqref{eq:H0} we have gauge transformed a finite bias into a time dependent tunneling matrix element, hence both chemical potentials are equal).

We then obtain
\begin{widetext}
\begin{subequations}
\begin{align}
\Re[I] &=- \sum_{\v x, \v x'} \Big \lbrace 2 t_{\v x}t_{\v x'}  \Im \left [\Pi^+_{1^\dagger 2}(\v x, \v x'; \mu_{21}) - 1 \leftrightarrow  2 \right ]\notag\\
&- 2 \int \frac{d\omega}{2\pi} {J_{\v x} J_{\v x'}} \Big \lbrace \Im[C^+(\v x, \v x'; \mu_{21}- \omega)] \Im [\Pi^+_{1^\dagger 2}(\v x, \v x'; \omega)] \left [\coth\left (\frac{\mu_{21}-  \omega}{2T}\right) + \coth\left (\frac{\omega}{2T}\right)\right ] - 1 \leftrightarrow  2 \Big \rbrace \Big \rbrace\notag\\
&=- \sum_{\v x, \v x'} \Big \lbrace 2 t_{\v x}t_{\v x'}  \Im \left [\Pi^+_{1^\dagger 2}(\v x, \v x'; \mu_{21}) - 1 \leftrightarrow  2 \right ]\notag\\
&- 2 \int_0^{\mu_{21}} \frac{d\omega}{2\pi} {J_{\v x} J_{\v x'}} \Im[C^+(\v x, \v x';\omega)] \Big \lbrace\Im [\Pi^+_{1^\dagger 2}(\v x, \v x'; \mu_{\rm 21}-\omega)]-\Im [\Pi^+_{2^\dagger 1}(\v x, \v x'; \omega - \mu_{\rm 21})] \Big \rbrace \Big \rbrace, \\
\Re[G(\Omega)] &= - \sum_{\v x, \v x'} \Big \lbrace \frac{2 t_{\v x} t_{\v x'}}{\Omega} \Im \left [\Pi^+_{1^\dagger 2}(\v x', \v x; \Omega) + 1 \leftrightarrow  2 \right ]\notag \\
	&-2 \int \frac{d\omega}{2\pi} \frac{J_{\v x} J_{\v x'}}{\Omega} \Big \lbrace \Im[C^+(\v x, \v x'; \Omega - \omega)] \Im[\Pi^+_{1^\dagger 2}(\v x, \v x'; \omega)]\left [\coth\left (\frac{\Omega-  \omega}{2T}\right) + \coth\left (\frac{\omega}{2T}\right)\right ] + 1 \leftrightarrow  2 \Big \rbrace \Big \rbrace \notag \\
&=- \sum_{\v x, \v x'} \Big \lbrace \frac{2 t_{\v x} t_{\v x'}}{\Omega} \Im \left [\Pi^+_{1^\dagger 2}(\v x, \v x'; \Omega) + 1 \leftrightarrow  2 \right ]\notag \\
&-2 \int_0^\Omega \frac{d\omega}{2\pi} \frac{J_{\v x} J_{\v x'}}{\Omega} \Big \lbrace \Im[C^+(\v x, \v x'; \omega)] \Im[\Pi^+_{1^\dagger 2}(\v x, \v x'; \Omega -\omega)] + 1 \leftrightarrow  2 \Big \rbrace \Big \rbrace.
\end{align}
\label{eq:ResponseApp}
\end{subequations}
\end{widetext}

In the second line, we assumed $\mu_{21} >0$ and $\Im[C^+(\v x, \v x'; \omega)] = -\Im[C^+(\v x', \v x; -\omega)]$, and also evaluated the distribution functions at zero temperature.
As the imaginary part of the particle-hole correlators is odd in frequency, it is evident that $I/V$ (as a function of V) and $G(\Omega)$ have the same functional form.	

This equation is the origin of Eq.~\eqref{eq:IMainText} of the main text, where we use the notation $\mathcal A^{\rm spin}_{\v x, \v x'}(\omega) = - 2 \Im C^{+}(\v x, \v x'; \omega)$, $\mathcal A^{1^\dagger 2}_{\v x, \v x'}(\omega) = - 2 \Im \Pi_{\rm 1^\dagger 2}^{+}(\v x, \v x'; \omega)$.

\subsection{Evaluation of electronic correlators}
In the case of point contact tunneling (when $t_{\v x} = t_0 \delta_{\v x, 0},J_{\v x} = J_0 \delta_{\v x, 0}$) we obtain 
\begin{eqnarray}
\Im[\Pi_{1^\dagger 2}^+(0,0;\omega)]
&=& -\pi \int dE \nu_B(E +\omega) \nu_1(E) \notag\\&&[f (E)  - f (E + \omega) ] \notag\\
&\simeq& -\pi \nu_B \nu_1 \omega.
\label{eq:ImPi}
\end{eqnarray}
Here, we used the assumption that the density of states $\nu_{1,2}(E)$ of the leads is only weakly energy dependent on the scale of the important energy scale $K$, justifying the notation $\nu_{1,2} = \nu_{1,2}(\mu_{1,2})$. If the point contact is between 2D systems $\nu_{1,2} = \nu_{\rm 2D} a^2$, where $a$ is the lattice constant and $\nu_{\rm 2D} = m/(2\pi)$.

For the case of planar tunneling we need the Fourier transform of Eq.~\eqref{eq:PiAs} 
\begin{subequations}
\begin{align}
\Im[\Pi_{1^\dagger 2}^+(\v q,\omega)] &= - \frac{1}{2} \int (dP) A_2(\epsilon, \v p) A_1(\epsilon - \omega, \v p - \v q) \notag \\ &[f(\epsilon - \omega) - f(\epsilon) ] \notag \\
&= - \nu_{\rm 2D} \sqrt{\frac{m  }{2q^2}} \Big \lbrace \sqrt{E_F - \left (\frac{m \omega^2}{2q^2} + \frac{q^2}{8m}\right) + \frac{\omega}{2} }\notag\\
&- \omega \rightarrow - \omega \Big\rbrace \label{eq:pH2D} \\
&\stackrel{\omega, v_F q \ll E_F}{\simeq} - \frac{\nu_{2D} \omega}{ \sqrt{(v_F q)^2 - \omega^2}}
\end{align}
\end{subequations}
Here, $v_F$ is the Fermi velocity and $E_F$ the Fermi energy and positivity of all arguments inside the square root is assumed (otherwise the contribution vanishes). In the presence of a finite lifetime, the low-energy limit is 
\begin{equation}
    \Im[\Pi_{1^\dagger 2}^+(\v q,\omega)] \simeq - \Re \frac{\nu_{2D} \omega}{ \sqrt{(v_F q)^2 - (\omega+ i \Gamma)^2}}.
\end{equation}
For a 1D tunneling barrier between 2D metals, we need
\begin{align}
\int (dq_x) \Im[\Pi_{1^\dagger 2}^+(q_x,q_y = 0,\omega)] \simeq  - \frac{\nu_{\rm 2D} \omega}{4 \pi v_F} \ln \left (\frac{4 E_F^2}{\omega^2 + \Gamma^2}\right).
\end{align}

Finally, we consider the interface of counter-propagating quantum Hall edges 
\begin{align}
\Im[\Pi_{1^\dagger 2}^+(x,x';\omega)] 
&=- \pi \nu_1 \nu_2 \int dE e^{i (2E + \omega) (x - x')/v_F} \notag \\
&[f(E) - f(E + \omega)] \notag \\
&=- \pi \nu_1 \nu_2\frac{v_F}{(x - x')} {\sin\left (\frac{\omega (x - x')}{v_F} \right)}.
\end{align}

\subsection{Current responses}

Using the above expressions, we first list the response for the elastic contributions (we here define $G_0 = 4\pi \nu_1 \nu_2 t_0^2 e^2/\hbar$, the tunneling conductance)
\begin{subequations}
\begin{align}
I_{\rm 0D}^{\rm el} &=G_0 V,\\
I_{\rm 1D}^{\rm el} &=G_0 (L/a) V \frac{1}{\pi v_F a \nu_{\rm 2D}} \ln \left ( \frac{4 E_F^2}{(eV)^2 + \Gamma^2} \right),\\
I_{\rm QH}^{\rm el} &=G_0 (L/a)\frac
{\pi v_F}{e a} \text{sign}(V),\\
I_{\rm 2D}^{\rm el} &=G_0 (L/a)^2 V \frac{1}{\pi\nu_{\rm 2D}a^2} \frac{\Gamma}{(eV)^2 + \Gamma^2}. 
\end{align}
\end{subequations}
For the quantum Hall setup, we used $\int_{-\infty}^\infty dx \sin(x)/x = \pi$, which leads to the sign function which is broadened on the scale $eV \sim v_F/L$. We emphasize that in this case, the given expression for the prefactor of the sign-function assumes  $L \ll v_F/t_0$ (it is obviously bound by the quantum of conductance)~\cite{BoeseZuelicke2001}. In the plots of these quantities, we use $4\pi v_F a \nu_{\rm 2D} = 2 p_F a$, $\nu_{2D} a^2 = (p_F a)^2/(4\pi E_F)$ with $p_F a= 1$, $E_F/K = 2.5, \Gamma/K =1/1000$. 

For the inelastic contribution in the case of the Kitaev material, one may use that the spin-correlator is short ranged on the scale of the leads. Thus, in any of the setups
\begin{equation}
I^{\rm inel} = -G_0  \int_0^{eV} (d\omega) (V -\omega/e) \sum_{\v x, \v x'} \frac{J_{\v x} J_{\v x'}}{t_0^2}\Im [C^+(\v x, \v x'; \omega)],
\end{equation}
i.e. 
\begin{subequations}
\begin{align}
\frac{d I_{\rm 0D}^{\rm inel}}{d V} &=-G_0 \frac{J_0^2}{t_0^2} \int_0^{eV} (d\omega) \Im [C^+(\v x, \v x; \omega)] ,\\
\frac{d I_{\rm 1D}^{\rm inel}}{d V}  &=-G_0 (L/a) \frac{J_0^2}{t_0^2} \int_0^{eV} (d\omega) \sum_{\v x'}\Im [C^+(\v x, \v x'; \omega)] ,\\
\frac{d I_{\rm QH}^{\rm inel}}{d V} &=-G_0 (L/a) \frac{J_0^2}{t_0^2} \int_0^{eV} (d\omega) \sum_{\v x'}\Im [C^+(\v x, \v x'; \omega)] ,\\
\frac{d I_{\rm 2D}^{\rm inel}}{d V} &= -G_0 (L/a)^2 \frac{J_0^2}{t_0^2} \int_0^{eV} (d\omega) \sum_{\v x'}\Im [C^+(\v x, \v x'; \omega)]. 
\end{align}
\end{subequations}

\section{Kitaev quantum spin liquid}
\label{sec:KitaevModel}

In this appendix we review some aspects of the Kitaev model and the spin-correlator for this particular QSL. Whilst trying to keep our presentation self-contained, several details are left out and can be found in original works such as Refs.~\cite{Kitaev2006,Baskaran2007,KnolleMoessner2014}. 

The main simplifying feature of Eq.~\eqref{eq:Kitaev} is an extensive number of conserved quantities (``plaquette operators''). This is also exploited in the exact solution by means of  fractionalization of spins into four Majorana operators $\hat S^\mu = i c^\mu c$. The local $\mathbb Z_2$ redundancy implies the emergence of a gauge theory, and as a matter of fact, the $c^\mu$ operators encode those gauge fields, see Fig.~\ref{fig:1}(d) of the main text. In the fractionalized approach, the conserved plaquette operators correspond to the flux through a given plaquette. Hence the model can be diagonalized for each flux configuration separately, the ground state displays a uniform flux solution (e.g. flux 0 through every plaquette). A single flipped plaquette (i.e. flux $\pi$) is called a vison. 

\subsection{Effective action of ``matter'' fields \& Green's function}

\begin{figure}
\includegraphics[width=.48\textwidth]{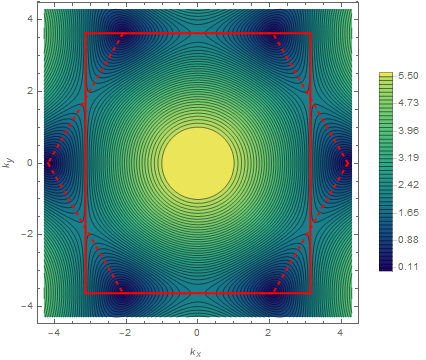}
\caption{Spectrum of spinon excitations (in units of $K$). The standard hexagonal Brillouin zone is represented dashed, while for the calculation of the on-site Green's function the rectangle (solid) is more convenient.}
\label{fig:Spectrum}
\end{figure}

In the flux free sector, the Hamiltonian is
\begin{equation}
H = \frac{1}{2} \int_{\rm BZ/2} (dk) (c^A, c^B)_{\v k} \left (\begin{array}{cc}
0 & 2 i s(\v k) \\ 
-2 i s^*(\v k) & 0
\end{array}  \right) \left (\begin{array}{c}
c^A \\ 
c^B
\end{array}  \right)_{\v k}.
\end{equation}
Here, $c^{A,B}_{\v k} = \sum_{\v x} e^{-i \v k \cdot \v x} c^{A,B}(\v x) = (c^{A,B}_{-\v k})^\dagger$ are Majorana operators on the A/B sublattice site of a given unit cell and $s(\v k) = K_x e^{i \v k \cdot \v n_1} + K_y e^{i \v k \cdot \v n_2} + K_z$ (for us $K_x =K_y = K_z = K$) as defined by Knolle and Moessner~\cite{KnolleMoessner2014}, whom we also follow in introducing $f(\v x) = [c^A(\v x) + i c^B(\v x)]/2$, so that
\begin{align}
H &=  \int_{\rm BZ} (dk) (f^\dagger_{\v k}, f_{- \v k}) \underbrace{\left (0, -\Im[s](\v k), \Re[s](\v k) \right )}_{=: \vec s(\v k)}\cdot \vec \tau \left (\begin{array}{c}
f_{\v k} \\ 
f_{-\v k}^\dagger
\end{array}  \right). \label{eq:NamubRepresentationHam}
\end{align}
Here, $\tau_{x,y,z}$ are matrices in sublattice space, which in this notation resembles the Nambu space of a spinless superconductor. 
The spectrum of excitations on top of the groundstate is given by $2 \vert s(\v k ) \vert \simeq \sqrt{3} K p$, where we linearized near the Dirac point $\v K = (2\pi/3 + p_x , 2\pi/\sqrt{3} p_y)$, see Fig.~\ref{fig:Spectrum}. 

The (Nambu) Green's function of $f$ fermions is 
\begin{equation}
G_{f}(z, \v k)  = [z - 2\vec s(\v k) \cdot \vec \tau]^{-1}.
\end{equation}
In particular, the onsite Green's function in the isotropic limit is
\begin{eqnarray}
G_f(z) &=& \int_{\rm BZ} (dk) \frac{z + 2\vec s(\v k)\cdot \vec \tau}{z^2 - 4\vert \vec s(\v k) \vert^2} \notag \\
&=& \frac{1}{4 K \sqrt{3}} \Big [ \frac{\bar z g(t)\tilde K(k(t))}{2\pi} \notag \\
&+& 2 \tau_z \left (  \frac{g(t)\tilde K(k(t))}{2\pi} \left (1+\frac{2t}{3} \right) - \frac{2}{3} \right) \Big ]. \label{eq:GreensFunction}
\end{eqnarray}
Here, ${\bar z = z/K; t = \frac{(\bar z/2)^2 - 3}{2}}$ and we used the result of Horiguchi~\cite{Horiguchi1972} 
which involves
\begin{subequations}
\begin{align}
g(t)&=[\sqrt{2t + 3}-1]^{-3/2}[\sqrt{2t + 3}+3]^{-1/2},  \\
k(t)&=(2t + 3)^{1/4} g(t)/2,  \\
\tilde K(k) &= K(k)  + 2 i \text{sign}(\Im(t)) K(\sqrt{1- k^2}) \Theta[\Im(k)\Im(t)] .
\end{align}
\end{subequations}
Here, $K(k) = \int_0^\pi [1- k^2 \sin^2(\theta)]^{-1/2} d\theta$.


\subsection{Spin correlators}

The greater spin-correlator is
\begin{align}
\tilde C_{ss'}^> (t_1,t_2) &= - i \langle S^{z}_s(t_1)S^z_{s'}(t_2)\rangle \notag\\
&= - i \langle e^{i H t_1} c^s c^{s,z} e^{-i H (t_1-t_2)} c^{s',z} c^{s'} e^{- i H t_2} \rangle .
\end{align}

By commuting the $z$ Majorana across the $e^{- i H(t_1 - t_2)}$ evolution operator, we generates a flip of the $z$ link. We further use ${c^{s,z}}^2 = 1$ and $ic^{A,z}c^{B,z} = 1$ when acting on the ground state. In the following, we express the expectation value only with respect to the ground state of matter fields $c^s$, only.
It is convenient to replace $c^{A,B}$ by $f, f^\dagger$ and use $\hat S(t_1,t_2) =  e^{i H t_1}e^{-i (H + V) (t_1-t_2)} e^{-i H t_2}$ (for $t_1>t_2:$ $\hat S(t_1,t_2) = T e^{- i \int_{t_1}^{t_2} d\tau V(\tau)}$) as well as $\tilde C_{AB} = \tilde C_{BA}$ and $\tilde C_{AA} = \tilde C_{BB}$ by crystalline symmetries. Here we used $V = -2K  \int_{\rm BZ} (dk) (f^\dagger_{\v k}, f_{- \v k}) \tau_z \left (\begin{array}{c}
f_{\v k} \\ 
f_{-\v k}^\dagger
\end{array}  \right)$, in addition to the unperturbed Hamiltonian introduced in Eq.~\eqref{eq:NamubRepresentationHam}.
Thus we obtain for $t_1 >t_2$
\begin{eqnarray}
\tilde C^>_{AB}(t_1,t_2) &=& \text{tr} [\mathbf G_f^q(t_1,t_2)\tau_z]/2\notag\\
\tilde C^>_{AA}(t_1,t_2) &=& \text{tr} [\mathbf G_f^q(t_1,t_2)]/2.
\end{eqnarray}
Here we use the Nambu formalism to express the same results as reported in the literature~\cite{KnolleMoessner2014,carrega2020tunneling} and
\begin{align}
\mathbf G_f^q(t_1,t_2) &= - i \left \langle  \left (\begin{array}{c} f(t_1) \\ f^\dagger(t_1)
\end{array}  \right) \hat S(t_1, t_2) \left (
f^\dagger(t_2) , 
f(t_2) \right) \right \rangle. \label{eq:QuenchCorrelator}
\end{align}

\subsection{Quench problem and approximate solution}


%
%
%


%
%
%

The correlator introduced in Eq.~\eqref{eq:QuenchCorrelator} contains a quench problem, as the flipped link abruptly appears at time $t_2$ and disappears at time $t_1$.
It is convenient to split off connected and disconnected part of the Green's function
\begin{eqnarray}
\underbrace{\mathbf G^q_f(t_1, t_2)}_{\text{``quench G.F.''}} &=& \underbrace{- i \frac{\left \langle  \left (\begin{array}{c} f(t_1) \\ f^\dagger(t_1)
\end{array}  \right) \hat S(t_1, t_2) \left (
f^\dagger(t_2) , 
f(t_2) \right) \right \rangle_0}{\left \langle  \hat S(t_1, t_2) \right \rangle_0}}_{=:\tilde{\mathbf{G}}_f(t_1, t_2) \text{ (``connected transient G.F.'')}}\notag \\
&&\times \underbrace{ \left \langle  \hat S(t_1, t_2) \right \rangle_0}_{=:e^{C(t_1,t_2)} \text{ (``loops'')}}.
\end{eqnarray} 

Since the Kitaev spin-liquid has semimetallic touching points (rather than a Fermi surface) there is no x-ray catastrophe. The bare Green's functions decay rapidly in time and thus all time integrals in the expansion of the the connected Green's function can be extended to infinity (a procedure called ``adiabatic approximation'' in Ref.~\cite{KnolleMoessner2014}). The connected Green's function is then easily resummed in frequency space (e.g. for the retarded Green's function)
\begin{equation}
    \tilde{\mathbf{G}}_f^+(\omega) \simeq \frac{G_f^+(\omega)}{1 + 4 K \tau_z G_f^+(\omega)}.
\end{equation}
Within this approximation, one may replace the ``loops'' contribution by $C(t_1 - t_2) = -i (t_1 - t_2)\Delta$ where $\Delta \simeq 0.26 K$\cite{Kitaev2006} is the energy difference between the ground state in the presence of zero (two adjacent) visons.
Then, the imaginary parts of the spin correlators at positive frequency are
\begin{equation}
\Im \left \lbrace \begin{array}{c} \tilde C^+_{AA}(\omega)\\ \tilde C^+_{AB}(\omega) \end{array} \right \rbrace =  \frac{\theta( \omega  - \Delta)}{2} \text{tr} \left [ \Im \tilde{\mathbf G}_f^+(\omega- \Delta)   \left \lbrace \begin{array}{c}\mathbf 1 \\ \tau_z \end{array} \right \rbrace \right], \label{eq:SpinCorrelator}
\end{equation}
which are plotted in Fig.~\ref{fig:SpectraAndDiDV}. In the main text of the paper we use the notation $C^+(\v x, \v x'; \omega) = 3 \tilde C^+_{AA}(\omega) \delta_{\v x, \v x'} + \tilde C^+_{AB}(\omega) \delta_{\langle \v x, \v x' \rangle}$.

\section{2D quantum antiferromagnet}
\label{sec:AFM}

In this section we summarize the calculations for the magnon contribution in a 2D quantum antiferromagnet (AFM) on a hexagonal lattice which underly Fig.~\ref{fig:AFM}. Our calculations are widely in parallel to the textbook~\cite{AuerbachBook}. 

\subsection{Hamiltonian in the large $S$ limit}

We employ a Holstein-Primakoff representation of spin-S operators.

\begin{eqnarray}
S_+ &=& \sqrt{2S - n_b} b, \\
S_- &=& b^\dagger \sqrt{2S - n_b}, \\
S_z &=& S - n_b.
\end{eqnarray}

In this convention, $\ket{m_z = S}= \ket{n_b = 0}$ and $\ket{m_z = -S}= \ket{n_b = 2S}$.
For the AFM with nearest neighbor interactions on a bipartite lattice, it is useful to rotate $S_x \rightarrow S_x, S_{y,z} \rightarrow - S_{y,z}$ on every other site. Then
\begin{align}
H &= J \sum_{<i,j>} \vec S_i \cdot \vec S_j \notag \\
& \rightarrow  -J \sum_{<i,j>} S_i^z S_j^z + \frac{J}{2} \sum_{<i,j>} \left (S_i^+ S_j^+ + S_i^- S_j^- \right) \notag \\
& \simeq  - \text{Vol}\, z J S^2 + Js \sum_{<i,j>} (2\hat n_i + b_i b_j + b_i^\dagger b_j^\dagger) + \mathcal O(1).
\end{align}
We use the same unit cell as in Sec.~\ref{sec:KitaevModel}, introduce $B^\dagger = (b_A^\dagger, b_B^\dagger)$ and drop the constant
\begin{equation}
H = J S z \int_{\rm BZ} (dk) (B_{\v k}^\dagger, B_{-\v k}^T) \left (\begin{array}{cc}
\mathbf 1 & h(\v k) \\ 
h(-\v k)^T & \mathbf 1
\end{array} \right ) \left (\begin{array}{c}
B_{\v k} \\ 
B_{- \v k}^*
\end{array}  \right).
\end{equation}
(We use the notation $B^* = (B^\dagger)^T$). The Hamiltonian
\begin{equation}
h(\v k) = \left (\begin{array}{cc}
0 & 1 + e^{- i \v k \cdot \hat n_1}+ e^{- i \v k \cdot \hat n_2} \\ 
1 + e^{ i \v k \cdot \hat n_1}+ e^{ i \v k \cdot \hat n_2} & 0
\end{array} \right)/3
\end{equation}
has the property $h(\v k) = h^T(- \v k)$ (i.e.~time reversal symmetry), eigenvalues $E_{\pm, \v k} \gtrless 0$ and corresponding eigenvectors $\ket{\psi_{\pm}(\v k)} = \ket{\psi_{\pm}(-\v k)}^*$.
We expand $B_{\v k} = \sum_{\pm} B_{\pm,\v k} \ket{\psi_{\pm}(\v k)}$. Then 
\begin{equation}
H =J S z\sum_\pm \int_{\rm BZ} (dk) (B_{\pm,\v k}^\dagger, B_{\pm,-\v k}) \left (\begin{array}{cc}
 1 & E_{\pm, \v k} \\ 
E_{\pm, \v k} & 1
\end{array} \right ) \left (\begin{array}{c}
B_{\pm,\v k} \\ 
B_{\pm,-\v k}^\dagger 
\end{array}  \right).
\end{equation}

We now use the Bogoliubov transform
\begin{eqnarray}
a_{\pm,\v k} &=& \cosh(\theta_{\v k}^\pm) b_{\pm,\v k} - \sinh(\theta_{\v k}^\pm) b_{\pm,-\v k}^\dagger, \\
b_{\pm,\v k} &=& \cosh(\theta_{\v k}^\pm) a_{\pm,\v k} + \sinh(\theta_{\v k}^\pm) a_{\pm,-\v k}^\dagger.
\end{eqnarray}

where $\tanh(\theta_{\v k}^\pm) = - E_{\pm, \v k}$. In this notation
\begin{equation}
H = 2 JS z\int (dk) (a^\dagger_{\pm,\v k}a_{\pm,\v k} + 1/2) \sqrt{1 - E_{\pm, \v k}^2}.
\end{equation}

\subsection{Spin correlator}

We consider the imaginary time ordered correlator (the time ordering operator is omitted for simplicity)
\begin{eqnarray}
C_{ij}(\tau, \tau') &=& -\langle \vec S_i (\tau) \cdot \vec S_j(\tau')\rangle \notag \\
&\rightarrow & - \eta_i \eta_j  \langle S_i^z (\tau) S_j^z(\tau')\rangle  \notag\\
&-&\frac{1 - \eta_i \eta_j}{4} \left ( \langle S_i^+ (\tau) S_j^+(\tau')\rangle +  \langle S_i^- (\tau) S_j^-(\tau')\rangle  \right) \notag \\
&-& \frac{1 + \eta_i \eta_j}{4} \left ( \langle S_i^+ (\tau) S_j^-(\tau')\rangle +  \langle S_i^- (\tau) S_j^+(\tau')\rangle  \right) \notag \\
&\simeq & -\eta_i \eta_j (S^2 - 2S(n_i + n_j)) \notag \\
&-&\frac{1 - \eta_i \eta_j}{2}S \left ( \langle b_i (\tau) b_j(\tau')\rangle +  \langle b_i^\dagger (\tau) b_j^\dagger(\tau')\rangle  \right) \notag \\
&-& \frac{1 + \eta_i \eta_j}{2}S \left ( \langle b_i (\tau) b_j^\dagger(\tau')\rangle +  \langle b_i^\dagger (\tau) b_j(\tau')\rangle  \right).\notag \\ &&
\end{eqnarray}
Here, $\eta_{i} = 1$ ($\eta_{i} = -1$) on the A (B) sublattice.

For our purposes, where electronic correlators are smooth on the lattice scale of the magnet, we need the average of same-sublattice and adjacent nearest sublattice correlators, c.f.~Eq.~\eqref{eq:ResponseApp},
\begin{align}
\sum_{i,j} \Pi_{1^\dagger 2}(i, j; \omega) C(i, j; \nu) &\simeq \sum_{\v x, \v x'} \Pi_{1^\dagger 2}(\v x, \v x'; \omega)  C_{\rm avg.}(\v x, \v x'; \nu), 
\end{align}
where ${i,j}$ are sites on the hexagonal lattice, while $\v x, \v x'$ are sites on the triangular Bravais lattice and
\begin{align}
C^+_{\rm avg.}(\v x, \v x'; \nu) &\simeq C^+_{AA}(\v x, \v x'; \nu) + C^+_{BB}(\v x, \v x'; \nu)  \notag\\
&+ \frac{1}{3} \sum_{\hat e = 0, \vec n_1, \vec n_2} [C^+_{AB}(\v x, \v x' + \hat e; \nu)\notag \\ 
&+ C^+_{BA}(\v x + \hat e, \v x'; \nu)].
\end{align}
The intrasublattice correlators are
\begin{equation}
C^+_{AA}(\v q; \omega) + C^+_{BB}(\v q;  \omega) =  2S \frac{G^+(\v q, \omega) + G^-(-\v q, -\omega)}{\sqrt{1 - E^2(\v q)}}.
\end{equation}
where 
$G^\pm(\v q, \omega) = [\omega\pm i \eta - 2JSz \sqrt{1 - E(\v q)^2}]$ and $E(\v q) =E_{+, \v q}$. Note that the sign $\xi = \pm 1$ associated to the two bands $E_{\xi, \v k}$ has dropped out of the equation.

Next we consider the sum of nearest neighbor terms, 
\begin{align}
&\frac{1}{3}\sum_{\hat e = 0, \vec n_1, \vec n_2}C_{AB}^+(\v x, \v x' - \hat e; \omega) + C_{BA}^+(\v x- \hat e, \v x' ;\omega)\notag \\
&= -2S \int_{\v q} e^{i\v q \cdot (\v x - \v x')} \frac{E(\v q)^2}{\sqrt{1-E(\v q)^2}} [G^+({\v q}, \omega) + G^-({-\v q}, -\omega)].
\end{align}

Note that the intersublattice correlator has opposite sign than the intrasublattice correlator for a N\'eel state: For example, if a spin-wave cants a spin in the direction of positive $\hat x$ at a given $A$ site, at the adjacent $B$ sites the spins are canted in the opposite direction (-$\hat x$) direction. Moreover, a $\v q$ dependent prefactor appears, which accounts for the details of the spatial dependence (clearly, for homogeneous $\v q = 0$ spin waves this prefactor is $E(\v q = 0)^2 = 1$, which corresponds to perfectly opposite correlators).

Thus we find for the total (averaged) retarded correlator
\begin{equation}
C^+_{\rm avg.}(\v q, \omega) =  2S {\sqrt{1 - E^2(\v q)}}[G^+(\v q, \omega) + G^-(-\v q, -\omega)].
\end{equation}

In particular $\Im[C^+_{\rm avg.}(\v q, \omega)] = -2\pi S {\sqrt{1 - E^2(\v q)}} \delta (\omega - 2JSz \sqrt{1 - E^2(\v q)})$ (for $\omega >0$),
where 
\begin{align}
&E(\v p) = \frac{1}{3}\sqrt{4 \cos \left(\frac{{p_x}}{2}\right) \cos \left(\frac{\sqrt{3} {p_y}}{2}\right)+2 \cos ({p_x})+3} \notag \\
&\Rightarrow \sqrt{1 - E^2(\v q)}) \simeq  \frac{q}{\sqrt{6}}, 
\end{align}
so that
\begin{equation}
\Im[C^+_{\rm avg.}(\v q, \omega)] \stackrel{v_s q \ll 2JSz}{\simeq} -2\pi S q \delta (\omega - v_s q)/\sqrt{6}.
\end{equation}

For the plot in Fig.~\ref{fig:AFM}, we extrapolate this result to $S = 1/2$ (where it is not formally controlled).
The inelastic tunneling current is thus given by

\begin{align}
I^{\rm inel.} &= -2\pi J_0^2 \int_0^{eV} (d\omega) \int (dq)  q \delta(\omega - v_s q)/\sqrt{6} \notag \\ &\Im[\Pi^+_{1^\dagger 2}(\v q, eV - \omega) - \Pi^+_{2^\dagger 1}(\v q, \omega - eV)],
\end{align}
with particle-hole correlators as determined in Eq.~\eqref{eq:pH2D}. 
In the low-voltage limit we find
\begin{align}
I^{\rm inel.} &\sim  \frac{J_0^2}{v_s^3} \int_0^{eV} d \omega \omega^2 \Re \frac{eV - \omega}{\sqrt{(v_F \omega/v_s)^2 - (eV - \omega^+)^2}} \notag\\
& = \frac{J_0^2 (eV)^3}{v_s^3} \int_0^{1} d \bar \omega \bar \omega^2 \Re \frac{1 - \bar \omega}{\sqrt{(v_F \bar \omega/v_s)^2 - (1- \bar \omega^+)^2}} ,
\end{align}
i.e.~a cubic current-voltage characteristic.

 \end{document}